\documentclass[12pt,leqno]{article}
\usepackage{amssymb}
\usepackage[mathscr]{eucal}
\usepackage{amsmath,amssymb,latexsym,theorem,bbm}
\usepackage{amsmath,amssymb,latexsym,theorem}
\usepackage{color,url}
\usepackage{appendix}
\usepackage{graphicx}
\usepackage[section]{placeins}

\newcommand{\DD}{\mathbb{D}}

\newcommand{\NN}{\mathbb{N}}

\newcommand{\RR}{\mathbb{R}}

\newcommand{\ZZ}{\mathbb{Z}}

\newcommand{\tA}{\widetilde{A}}

\newcommand{\cB}{{\mathcal B}}

\newcommand{\cD}{{\mathcal D}}

\newcommand{\cN}{{\mathcal N}}

\newcommand{\dd}{\mathrm{d}}
\newcommand{\ee}{\mathrm{e}}

\newcommand{\EE}{\operatorname{\mathbb{E}}}

\newcommand{\PP}{\operatorname{\mathbb{P}}}

\newcommand{\VaR}{\operatorname{ \tt{VaR} } }

\newcommand{\ES}{\operatorname{ \tt{ES}} }
\newcommand{\DiffVaR}{\operatorname{ \tt{DiffVaR} } }
\newcommand{\DiffES}{\operatorname{ \tt{DiffES} } }

\newcommand{\tS}{\widetilde{S}}

\renewcommand{\mid}{\,|\,}

\renewcommand{\leq}{\leqslant}
\renewcommand{\geq}{\geqslant}

\newcommand{\distr}{\stackrel{\cD}{\longrightarrow}}

\newcommand{\bbone}{\mathbbm{1}}

\newcommand{\proofend}{\hfill\mbox{$\Box$}}

\numberwithin{equation}{section}

\theoremstyle{change} \theorembodyfont{\em}
\newtheorem{Lem}{Lemma.}[section]

\newtheorem{Pro}[Lem]{Proposition.}

\newtheorem{Def}[Lem]{Definition.}

\theorembodyfont{\rm}
\newtheorem{Rem}[Lem]{Remark.}
\newtheorem{Ex}[Lem]{Example.}

\begin{document}

\begin{center}
 {\bfseries\Large
    On approximations of Value at Risk and Expected Shortfall involving kurtosis}

\vspace*{3mm}

 {\sc\large
  M\'aty\'as $\text{Barczy}^{*,\diamond}$,
  \ \'Ad\'am $\text{Dud\'as}^{**}$,
  \ J\'ozsef $\text{G\'all}^{***}$}

\end{center}

\vskip0.2cm

\noindent
 * MTA-SZTE Analysis and Stochastics Research Group,
   Bolyai Institute, University of Szeged,
   Aradi v\'ertan\'uk tere 1, H--6720 Szeged, Hungary.

\noindent
 ** Former master student of Faculty of Science and Technology, University of Debrecen, Hungary.

\noindent
 *** Faculty of Informatics, University of Debrecen,
    Pf.~12, H--4010 Debrecen, Hungary.

\noindent e-mail: barczy@math.u-szeged.hu (M. Barczy),
                  adamdudas92@gmail.com (\'A. Dud\'as),\\
                  gall.jozsef@inf.unideb.hu (J. G\'all).

\noindent $\diamond$ Corresponding author.


\renewcommand{\thefootnote}{}
\vspace*{0.2cm}
\footnote{M\'aty\'as Barczy is supported by the J\'anos Bolyai Research Scholarship
 of the Hungarian Academy of Sciences.}

\vspace*{-10mm}

\begin{abstract}
We derive new approximations for the Value at Risk and the Expected Shortfall at high levels
 of loss distributions with positive skewness and excess kurtosis, and we describe their precisions for notable ones such as for exponential,
 Pareto type I, lognormal and compound (Poisson) distributions.
Our approximations are motivated by that kind of extensions of the so-called Normal Power Approximation,
 used for approximating the cumulative distribution function of a random variable, which incorporate
 not only the skewness but the kurtosis of the random variable in question as well.
We show the performance of our approximations in numerical examples and we also give comparisons with some known ones in the literature.
\end{abstract}

\section{Introduction}
\label{section_intro}

Value at Risk \ $(\VaR)$ \ and Expected Shortfall \ $(\ES)$ \ are standard risk measures
  in financial and insurance mathematics.
 \ $\VaR$ \ permits to measure the maximum aggregate loss of a portfolio with a given confidence level,
 while \ $\ES$ \ can be defined as the conditional expectation of the loss for losses beyond
 the corresponding $\VaR$-level (see Definitions \ref{Def_VaR} and \ref{Def_ES}).
Closed formulas for \ $\VaR$ \ and \ $\ES$ \ are mainly available for random variables
 with explicitly given density functions, for a comprehensive list with more than one hundred examples
 see Nadarajah et al.\ \cite{NadChaAfu}.
For compound (Poisson) distributions such closed formulas are rarely available, so the approximations of \ $\VaR$ \ and \ $\ES$ \ in these cases
 are of high importance.
There is a vast literature on the properties and the approximation of the distribution function of losses
 which are often used in the collective model of insurance mathematics or for calculating operational risk in finance.
To name some very recent works, we can mention Roozegar and Nadarajah \cite{RozNad} and  Bar-Lev and Ridder \cite{BarRid}
 on special collective risk models.
For a good survey on the comparison of approximations for the distribution function of a compound Poisson distribution,
 see Seri and Choirat \cite{SerCho}.

In this paper we derive new approximations for \ $\VaR$ \ and \ $\ES$ \ at high levels of a loss distribution having a continuous distribution
 function with positive skewness and excess kurtosis using its first four moments (provided that they are finite), see Section \ref{Section_VaR_ES}.
We study their precisions for notable loss distributions such as for exponential, Pareto type I, lognormal and compound (Poisson) distributions.
The reason for the popularity of exponential distribution is due to one's own convenience, rather than it can be well-fitted to loss data in general.
On the one hand, its parameter can be well-estimated using only the sample mean, and on the other hand, it is easy to work with it from
 the point of view making explicit calculations.
Lognormal distributions usually show much better fitting to loss data, but estimating their two parameters is more complicated.
Pareto distributions can be well-used for modelling big losses in insurance such as losses due to forest fire,
 since Pareto distributions are in the family of so-called heavy tailed distributions.
Compound (Poisson) distributions are widely used in modelling the total loss in a given time period in insurance and finance (e.g. for operational risk),
 so recursive and approximative formulae for its distribution function, $\VaR$ \ and \ $\ES$ \ are of high importance as well.
In fact, our approximations for \ $\VaR$ \ and \ $\ES$ \ can be formally used for any distribution function with finite fourth moment
 having positive skewness and excess kurtosis,
 though their behaviour and precisions should carefully be studied for every particular case.
Baixauli and Alvarez \cite{BaiAlv} showed empirical evidence that the kurtosis contributes to obtain more precise
 \ $\VaR$ approximations using data from seven stock indices such as S$\&$P500 and NIKKEI,
  which underlines the necessity of approximations of \ $\VaR$ \ and \ $\ES$ \ containing the kurtosis of the loss as well.

Our approximative formulas will be motivated by extensions of the so-called Normal Power Approximation (NPA).
For a random variable \ $S$ \ with \ $\EE(\vert S\vert^3)<\infty$, \ $\DD^2 (S)\ne 0$, \ and positive skewness \ $\gamma_S$ \ (recalled below),
 one can obtain the following approximation for the distribution function of the standardized version of \ $S$:
  \begin{align}\label{NPA}
    \PP\left( \frac{S-\EE(S)}{\sqrt{\DD^2(S)}} < x+ \frac{\gamma_S}{6}(x^2-1) \right) \approx \Phi(x),
    \qquad  x\in\RR,
 \end{align}
 where \ $\DD^2(S):=\EE( (S-\EE(S))^2 )$ \ and
 \[
  \gamma_S:=\frac{\EE((S- \EE(S))^3)}{ (\EE((S- \EE(S))^2) )^{3/2}}
           = \frac{\EE((S- \EE(S))^3)}{ (\DD^2(S))^{3/2}}
 \]
 is the skewness of \ $S$ \ and \ $\Phi$ \ denotes the distribution function
 of a standard normally distributed random variable.
Formula \eqref{NPA} is called the NPA for \ $F_S$, \ and it is usually credited to K. Loimaranta,
 see Kauppi and Ojantakanen \cite[page 219]{KauOja}.
For other references on NPA, see e.g. Beard et al.\ \cite[Section 3.11]{BeaPenPes}, Daykin et al.\ \cite[Section 4.2.4]{DayPenPes},
Kaas et al.\ \cite[Section 2.5.3]{KaaGooDhaDen} or Seri and Choirat \cite[Section 5]{SerCho}.
Here we call the attention to the fact that we used \ $\approx$ \ in formula \eqref{NPA} to indicate that
 the error of the approximation is not derived in a rigorous mathematical way in the literature in general.
According to our knowledge, the only special case when the asymptotic behaviour of the error was studied
 is the case of \ $S$ \ having a compound Poisson distribution due to Seri and Choirat \cite[Section 5]{SerCho}.
In all what follows \ $\approx$ \ will be meant in the same way.
We will use the extensions of \eqref{NPA} only for motivations for introducing new approximations of \ $\VaR$ \ and \ $\ES$ \ of \ $S$ \
 in Section \ref{Section_VaR_ES} of which the precisions will be carefully studied for particular loss distributions.
Daykin et al.\ \cite{DayPenPes} point out that, in practice, \eqref{NPA} is suggested to be used for the right-tail of \ $S$ \
 and so far as the skewness \ $\gamma_S$ \ does not exceed \ $1$ \ or at most \ $1.2$, \
 otherwise it becomes unreliable as they note.
Note also that if \ $x\geq 0$, \ then \eqref{NPA} can be rewritten in the form
 \begin{align}\label{NPA_atiras}
  \PP\left( \frac{S- \EE(S)}{\sqrt{\DD^2(S)}} < y \right)
      \approx \Phi\left( \sqrt{\frac{9}{\gamma_S^2}  +  \frac{6y}{\gamma_S} + 1} - \frac{3}{\gamma_S}\right),
      \qquad y\geq -\frac{\gamma_S}{6}.
 \end{align}

Motivated by \eqref{NPA}, one can introduce an approximation of \ $\VaR_S(\alpha)$ \ (the \ $\VaR$ \ of \ $S$ \ at a level \ $\alpha$) \ given by
 \begin{align}\label{VaR_approx_NPA}
    \EE(S)+\sqrt{\DD^2(S)}\Big( z_\alpha + \frac{\gamma_S}{6}(z_\alpha^2-1) \Big)
 \end{align}
 for any \ $\alpha\in(0,1)$, \ where \ $z_\alpha$ \ denotes the quantile of a standard normal distribution at a confidence level \ $\alpha$,
 \ see Casta\~{n}er et al.\ \cite[Lemma 2]{CasClaMar}.
Further, motivated by \eqref{VaR_approx_NPA}, one can introduce an approximation of \ $\ES_S(\alpha)$ \
 (the \ $\ES$ \ of \ $S$ \ at a level \ $\alpha$) \ given by
 \begin{align}\label{ES_approx_NPA}
    \EE(S)+\sqrt{\DD^2(S)} \frac{\varphi(z_\alpha)}{1-\alpha}\Big( 1 + \frac{\gamma_S}{6}z_\alpha \Big)
 \end{align}
 for any \ $\alpha\in(0,1)$, \ where \ $\varphi(x):=\frac{1}{\sqrt{2\pi}} \ee^{-\frac{x^2}{2}}$, $x\in\RR$,
 \ denotes the density function of a standard normally distributed random variable, see Casta\~{n}er et al.\ \cite[Theorem 1]{CasClaMar}.

In the literature, one can find a refinement of the NPA \eqref{NPA}
 which also involves the excess kurtosis of \ $S$ \ defined by
 \[
 \kappa_S:=\frac{\EE((S- \EE(S))^4)}{ (\EE((S- \EE(S))^2) )^2} - 3
           = \frac{\EE((S- \EE(S))^4)}{ (\DD^2(S))^2} - 3.
 \]
Namely, for a random variable \ $S$ \ with \ $\EE(S^4)<\infty$, \ $\DD^2 (S)\ne 0$, \ $\gamma_S>0$ \ and \ $\kappa_S>0$,
 \ one can obtain the following approximation for the distribution function of the standardized version of \ $S$:
 \begin{align}\label{ENPA_Beard_et_al}
  \PP\left( \frac{S-\EE(S)}{\sqrt{\DD^2(S)}} < x + \frac{\gamma_S}{6}(x^2-1) + \frac{\kappa_S}{24} (x^3 -3x)
                                                 - \frac{\gamma_S^2}{36} (2x^3 - 5x)\right) \approx \Phi(x),
    \qquad x\in\RR,
 \end{align}
 see Kauppi and Ojantakanen \cite[page 221]{KauOja}, Beard et al.\ \cite[(3.11.2) and (3.11.9)]{BeaPenPes}
 or Seri and Choirat \cite[Section 8]{SerCho}.
In case of \ $S$ \ has a compound Poisson distribution, the precision of the approximation \eqref{ENPA_Beard_et_al}
 is also given, see Seri and Choirat \cite[Section 8]{SerCho}, but, in general, we are not aware of such an analysis.
In fact, Seri and Choirat \cite[Section 8]{SerCho} presented another form of \eqref{ENPA_Beard_et_al},
 which can be considered as the counterpart of \eqref{NPA_atiras} involving \ $\kappa_S$ \ as well.
For a given \ $y\in\RR$, \ they considered the cubic equation
 \ $ x + \frac{\gamma_S}{6}(x^2-1) + \frac{\kappa_S}{24} (x^3 -3x)
 - \frac{\gamma_S^2}{36} (2x^3 - 5x) =y$, \ $x\in\RR$, \ which can have three solutions in general,
 and, for the other form of \eqref{ENPA_Beard_et_al} in question, they simply chose an appropriate root using a rule of thumb.
Seri and Choirat \cite[Section 19]{SerCho} also noted that the approximation \eqref{ENPA_Beard_et_al} is among the best four ones of the 15 approximation
 methods that they compared.

Motivated by \eqref{ENPA_Beard_et_al}, one can introduce an approximation of \ $\VaR_S(\alpha)$ \ given by
 \begin{align}\label{VaR_approx_CF}
  \EE(S)+\sqrt{\DD^2(S)}\Big( z_\alpha + \frac{\gamma_S}{6}(z_\alpha^2-1) + \frac{\kappa_S}{24} (z_\alpha^3 -3z_\alpha)
                                                 - \frac{\gamma_S^2}{36} (2z_\alpha^3 - 5z_\alpha)\Big)
 \end{align}
 for any \ $\alpha\in(0,1)$, \ known as the Cornish-Fisher's approximation of \ $\VaR_S(\alpha)$,
 \ see, e.g., Alexander \cite[formula (IV.3.7)]{Ale}.
Further, motivated by \eqref{VaR_approx_CF}, one can introduce an approximation of \ $\ES_S(\alpha)$ \ given by
 \begin{align}\label{ES_approx_CF}
    \EE(S)+\sqrt{\DD^2(S)} \frac{\varphi(z_\alpha)}{1-\alpha}\Big( 1 + \frac{\gamma_S}{6}z_\alpha + (z_\alpha^2 - 1)\frac{\kappa_S}{24}
                                                                     + (1-2z_\alpha^2)\frac{\gamma_S^2}{36} \Big)
 \end{align}
 for any \ $\alpha\in(0,1)$, \ see Maillard \cite[Section 6]{Mai}.

Recently, Lien et al.\ \cite{LieYanYe} recalled and compared some alternative approximations of \ $\VaR_S$ \ such as the Sillitto's
 approximation which is based on so-called $L$-moments.

In Section \ref{Section_ENPA}, we present other refinements of \eqref{NPA} which also involve the excess kurtosis
 \ $\kappa_S$ \ of \ $S$ \ such as
  \begin{align}\label{KNPA1}
  &\PP\left( \frac{S-\EE(S)}{\sqrt{\DD^2(S)}}
            < x + \frac{ - \frac{\gamma_S}{6}(x^2-1) + \frac{\kappa_S}{24} (-x^3+3x)}
                        { -1 + \frac{\gamma_S}{6}(-x^3+3x) - \frac{\kappa_S}{24} (x^4 - 6x^2 + 3)  } \right)
      \approx \Phi(x),
 \end{align}
 provided that \ $\gamma_S>0$, \ $\kappa_S>0$ \ and \ $-1 + \frac{\gamma_S}{6}(-x^3+3x) - \frac{\kappa_S}{24} (x^4 - 6x^2 + 3) \ne 0$.
\ Note that, if \ $\gamma_S>0$ \ and \ $\kappa_S>0$, \ then \ $ -1 + \frac{\gamma_S}{6}(-x^3+3x) - \frac{\kappa_S}{24} (x^4 - 6x^2 + 3) <0$
 \ for all \ $x>\sqrt{3+\sqrt{6}}\approx 2.334$, \ see also part (iii) of Remark \ref{Rem_VaR_region}.
In principle, one could ask for rewriting \eqref{KNPA1} in the same spirit as \eqref{NPA_atiras}.
For this, one should solve an equation of order four in general, and proceed as in Section 8 in
 Seri and Choirat \cite{SerCho}.
We do not deal with it, since we do not need such a version, and formula \eqref{KNPA1} will be suitable for our purposes
 to obtain our new approximations of \ $\VaR_S(\alpha)$ \ and \ $\ES_S(\alpha)$ \ at a high level
 of \ $\alpha\in(0,1)$ \ given in Section \ref{Section_VaR_ES}.
For other refinements of \eqref{NPA}, see \eqref{KNPA4}, \eqref{KNPA5} and \eqref{KNPA6}.
The derivation of \eqref{KNPA1} is a based on a so-called $4^{\mathrm{th}}$ order Gram-Charlier type A
 expansion of the distribution function \ $F_S$ \ of \ $S$ \ (for details, see Section \ref{Section_ENPA}),
 and in part (i) of Remark \ref{Rem_NR} we point out a connection between \eqref{KNPA1} and Newton-Raphson's recursion as well.
In Remark \ref{Rem_GC} we point out that in case of \ $\kappa_S>0$, \ the $4^{\mathrm{th}}$ order Gram-Charlier type A
 expansion of \ $F_S$ \ can be used as an approximation of \ $F_S$ \
 in the sense that it behaves as a distribution function for large enough \ $x\in\RR$,
 and from a practical point of view the condition \ $\kappa_S>0$ \ is not so restrictive, since
 \ $\kappa_S$ \ is positive for popular loss distributions such as for exponential, Pareto type I, lognormal
 or many compound (Poisson) distributions.
The formula \eqref{ENPA_Beard_et_al} due to Kauppi and Ojantakanen \cite{KauOja} and our formulae \eqref{KNPA1}, \eqref{KNPA4},
 \eqref{KNPA5} and \eqref{KNPA6} are in the same spirit.
However, there are some subtle details resulting different formulae.
On the one hand, the formula \eqref{ENPA_Beard_et_al} is based on a $4^{\mathrm{th}}$ order Edgeworth expansion of \ $F_S$,
 which results an additional term \ $\frac{\gamma_S^2}{72}\Phi^{(6)}(x)$ \ compared to the $4^{\mathrm{th}}$ order Gram-Charlier type A expansion
 of \ $F_S$ \ in \eqref{GC}.
On the other hand, formulae \eqref{KNPA1}, \eqref{KNPA4}, \eqref{KNPA5} and  \eqref{KNPA6} are derived
 using the Newton's approximation for a solution to a nonlinear equation \ $g_x(\delta)=0$ \ (see \eqref{help10_gx}).

In Section \ref{Section_VaR_ES}, motivated by the refinements \eqref{KNPA1}, \eqref{KNPA4},
 \eqref{KNPA5} and \eqref{KNPA6} of the NPA \eqref{NPA},
 we introduce new approximations of \ $\VaR_S(\alpha)$ \ and \ $\ES_S(\alpha)$ \ at a high level
 of \ $\alpha\in(0,1)$ \ (mostly above \ $0.99$) \ involving the excess kurtosis \ $\kappa_S$ \  of \ $S$ \ as well.
In case of the refinements \eqref{KNPA1} and \eqref{KNPA6}, the corresponding new approximations of the risk measures
 can be found in \eqref{VAR_approx_mod}, \eqref{VAR_approx_mod_IV} and \eqref{ES_approx_mod}, \eqref{ES_approx_mod_IV}, respectively.
Since the Basel Committee usually sets the confidence level at \ $0.999$ \ for $\VaR$ and at \ $0.975$ \ for $\ES$ \ (see, for instance, \cite[33.20/(5) and 33.3]{BasCom}),
 it is reasonable to focus on approximations of \ $\VaR_S(\alpha)$ \ and \ $\ES_S(\alpha)$ \ which behave well for $\alpha$ above $0.99$.
\ The approximations \eqref{VAR_approx_mod_IV} and \eqref{ES_approx_mod_IV} can be considered as modifications of
 \eqref{VaR_approx_NPA} and \eqref{ES_approx_NPA}, respectively, which contain only the first three moments of \ $S$,
  \ but it measures the effect of the skewness \ $\gamma_S$ \ in another way.
In Proposition \ref{Pro_VaR_ES_aszimptotika} we describe their asymptotic behaviour as \ $\alpha\uparrow 1$,
 \ and it turns out that they behave asymptotically exactly in the same way.
\ Further, we study the precisions of these approximations for notable loss distributions such as for exponential, Pareto type I, lognormal and compound (Poisson) distributions,
 see Examples \ref{Ex_Exp}, \ref{Ex_Par}, \ref{Ex_LN} and \ref{Ex_CP}.
In case of a compound Poisson distribution it turns out that for large enough \ $\alpha$, \ the difference
 of \ $\VaR_S(\alpha)$ \ and its approximated value normalized by the deviation of the compound Poisson distribution
 in question converges to 0 as the expected number of the claims tends to infinity, see \eqref{CP_precision}.
We also point out to the fact that our approximative formulae \eqref{VAR_approx_mod}, \eqref{ES_approx_mod}, \eqref{VAR_approx_mod_IV} and \eqref{ES_approx_mod_IV}
 are explicit in terms of the first four moments of \ $S$, \ so one can use them for a sensitivity analysis, i.e., one can study
 how these approximative formulae depend on some parameters of \ $S$.
\ In part (vi) of Remark \ref{Rem_VaR_region} we study the question how the approximative formulae \eqref{VAR_approx_mod} and \eqref{ES_approx_mod}
 are transformed provided that the underlying loss distribution \ $S$ \ is transformed by an affine transformation.

In Section \ref{Section_comparisons}, we show the performance of our new approximative formulae
 and compare them with those of some known (explicite) ones, namely, \eqref{VaR_approx_NPA},
 \eqref{VaR_approx_CF}, \eqref{ES_approx_NPA} and \eqref{ES_approx_CF} in case of the above mentioned notable loss distributions
 under different choices of parameters and at high levels of \ $\alpha$ \ (mostly above \ $0.99$).
\ For exponential, Pareto type I and lognormal distributions \ $S$, \ explicit formulae are available for the theoretical \ $\VaR_S(\alpha)$
 \ (see, e.g., Examples \ref{Ex_Exp}, \ref{Ex_Par} and \ref{Ex_LN}), but for a compound Poisson distribution no such explicit formula is available.
 We also give comparison of our approximative formulae for the compound (Poisson) case with the Monte-Carlo estimates.
Our studies shows that the new approximative formulae \eqref{VAR_approx_mod} and \eqref{VAR_approx_mod_IV} outperform
 the other known formulae in many numerical cases.
We also note that \eqref{VAR_approx_mod} can overestimate and underestimate \ $\VaR_S(\alpha)$, \ $\alpha\in(0.991,1)$, \ as well
 depending on the value of \ $\alpha$.
\ In risk assessment, underestimation is a much more serious issue (which might mean insolvency in capital calculation).
However, one should bear in mind that \eqref{VAR_approx_mod} can have zero relative error as well which
 shows an advantage of application of this approximation method.
For a more detailed discussion on our numerical results  and on the comparison of the methods,
  see Section \ref{Section_comparisons}.

\section{Extensions of Normal Power Approximation using kurtosis}
\label{Section_ENPA}

Let \ $\ZZ_+$, \ $\NN$ \ and \ $\RR$ \ denote the
 set of non-negative integers, positive integers and real numbers, respectively.
The Borel $\sigma$-algebra on \ $\RR$ \ will be denoted by \ $\cB(\RR)$.
For functions \ $f : D \to \RR$ \ and \ $g : D \to (0,\infty)$, \ where \ $D\subseteq \RR$,
 \ by the notations \ $f(x) \sim g(x)$ \ and \ $f(x) = o(g(x))$ \ as \ $x \to x_0$, \ with some \ $x_0\in\RR\cup\{\infty\}$,
 \ we mean that \ $\lim_{x\to x_0} \frac{f(x)}{g(x)} = 1$ \ and \ $\lim_{x\to x_0} \frac{f(x)}{g(x)} = 0$, \ respectively.
The notation \ $f(x) = O(g(x))$ \ as \ $\vert x\vert \to\infty$ \ means that there exist constants \ $C\in(0,\infty)$ \ and
 \ $x_C\in\RR$ \ such that \ $\vert f(x)\vert \leq C g(x)$ \ for all \ $\vert x\vert\geq x_C$.

Based on Kaas et al.\ \cite[Remark 2.5.9]{KaaGooDhaDen} and W\"uthrich \cite[Section 4.1.3]{Wut}
 we give extensions of the NPA \eqref{NPA} incorporating the excess kurtosis
 of the underlying random variable as well.

Let \ $S$ \ be a random variable such that \ $\DD^2 (S)\ne 0$ \ and its moment generating function is finite on an interval \ $(-t_0,t_0)$,
 \ where \ $t_0>0$, \ i.e., \ $M_S(t):=\EE(\ee^{t S})<\infty$ \ for each \ $t\in(-t_0,t_0)$.
Especially, \ $\EE(\vert S\vert^i)<\infty$, \ $i=1,2,3,4$.
\ Let \ $Z$ \ denote the standardized version of \ $S$, \ i.e., \ $Z:=(S-\EE(S))/\sqrt{\DD^2(S)}$.
\ Let \ $\gamma_Z:=\EE(Z^3)$ \ and \ $\kappa_Z:=\EE(Z^4) - 3$.
\ Note that \ $\EE(Z)=0$, \ $\EE(Z^2)=1$, \ and \ $\gamma_Z$ \ and \ $\kappa_Z$ \ are the skewness and excess kurtosis
 of \ $Z$ \ (and of \ $S$ \ as well, i.e., \ $\gamma_Z = \gamma_S$ \ and \ $\kappa_Z=\kappa_S$), \ respectively.
Further, the moment generating function of \ $Z$ \ is finite on \ $(-t_0,t_0)$ \ as well.

We will derive refinements of the NPA \eqref{NPA} which also involve the excess kurtosis
 \ $\kappa_S$ \ of \ $S$ \ such as
  \begin{align*}
  &\PP\left( \frac{S-\EE(S)}{\sqrt{\DD^2(S)}}
            < x + \frac{ - \frac{\gamma_S}{6}(x^2-1) + \frac{\kappa_S}{24} (-x^3+3x)}
                        { -1 + \frac{\gamma_S}{6}(-x^3+3x) - \frac{\kappa_S}{24} (x^4 - 6x^2 + 3)  } \right)
      \approx \Phi(x),
 \end{align*}
 provided that \ $\gamma_S>0$, \ $\kappa_S>0$ \ and \ $-1 + \frac{\gamma_S}{6}(-x^3+3x) - \frac{\kappa_S}{24} (x^4 - 6x^2 + 3) \ne 0$, \ $x\in\RR$.
 \ Note that, if \ $\gamma_S>0$ \ and \ $\kappa_S>0$, \ then \ $ -1 + \frac{\gamma_S}{6}(-x^3+3x) - \frac{\kappa_S}{24} (x^4 - 6x^2 + 3) <0$
 \ for all \ $x>\sqrt{3+\sqrt{6}}$, \ see also part (iii) of Remark \ref{Rem_VaR_region}.
\ For some other refinements of \eqref{NPA}, see \eqref{KNPA4}, \eqref{KNPA5} and \eqref{KNPA6}.

By a $4^{\mathrm{th}}$ order Taylor's expansion of the cumulant generating function \ $\log(M_Z(t))$, \ $t\in(-t_0,t_0)$ \ of \ $Z$,
 \ we have
 \[
    \log(M_Z(t)) = \sum_{k=0}^4 \frac{(\log(M_Z))^{(k)}(0)}{k!} t^k + o(t^4)
          \qquad \text{as \ $t\to 0$,}
 \]
 where
 \begin{align*}
   & (\log(M_Z))^{(0)}(0) = 0,\qquad (\log(M_Z))^{(1)}(0) = \EE(Z) =0, \qquad (\log(M_Z))^{(2)}(0) = \EE(Z^2) = 1,\\
   & (\log(M_Z))^{(3)}(0) = \EE(Z^3) = \gamma_Z, \qquad  (\log(M_Z))^{(4)}(0) = \EE(Z^4) - 3 = \kappa_Z.
 \end{align*}
For completeness, we presented the calculations for the first four cumulants
 \ $(\log(M_Z))^{(k)}(0)$, \ $k\in \{1,2,3,4\}$ \ of \ $Z$ \ in Appendix \ref{App_kumulans}.
Hence
 \begin{align*}
  \log(M_Z(t)) = \frac{1}{2}t^2 + \frac{\gamma_Z}{6}t^3 + \frac{\kappa_Z}{24}t^4 + o(t^4)
          \qquad \text{as \ $t\to 0$,}
 \end{align*}
 and
 \[
   M_Z(t) = \exp\left\{ \frac{1}{2}t^2 + \frac{\gamma_Z}{6}t^3 + \frac{\kappa_Z}{24}t^4 + o(t^4)\right\}
          = \ee^{\frac{t^2}{2}} \exp\left\{\frac{\gamma_Z}{6}t^3 + \frac{\kappa_Z}{24}t^4 + o(t^4)\right\}
          \qquad \text{as \ $t\to 0$.}
 \]
Using a first order Taylor's approximation of the second exponential function in the formula above (i.e.,
  \ $\ee^x = 1+ x + o(x)$ \ as \ $x\to0$) \ we have
 \begin{align}\label{help1}
    M_Z(t) \sim  \ee^{\frac{t^2}{2}} + \frac{\gamma_Z}{6}t^3 \ee^{\frac{t^2}{2}} + \frac{\kappa_Z}{24}t^4 \ee^{\frac{t^2}{2}}
                   + o(t^3)
              \qquad \text{as \ $t\to 0$.}
 \end{align}
Indeed, if \ $f:\RR\to\RR$ \ is a function such that \ $f(t) = o(t^4)$ \ as \ $t\to0$, \ then
 \begin{align*}
              &\lim_{t\to 0} \frac{\Big \vert \ee^{\frac{t^2}{2}}\exp\left\{ \frac{\gamma_Z}{6}t^3 + \frac{\kappa_Z}{24}t^4  + f(t) \right\}
                       -  \big(\ee^{\frac{t^2}{2}} + \frac{\gamma_Z}{6}t^3 \ee^{\frac{t^2}{2}} + \frac{\kappa_Z}{24}t^4 \ee^{\frac{t^2}{2}} \big) \Big \vert }
                            {t^3} \\
              &= \lim_{t\to 0} \frac{\ee^{\frac{t^2}{2}}}{t^3}
                    \left\vert f(t) + \sum_{k=2}^\infty \frac{ \big(\frac{\gamma_Z}{6}t^3 + \frac{\kappa_Z}{24}t^4  + f(t) \big)^k}{k!} \right\vert
                \leq  \lim_{t\to0} \frac{\ee^{\frac{t^2}{2}}}{t^3}
                \sum_{k=2}^\infty \frac{ \big\vert \frac{\gamma_Z}{6}t^3 + \frac{\kappa_Z}{24}t^4 + f(t)\big\vert^k}{k!} \\
              & \leq \lim_{t\to 0}{\ee^{\frac{t^2}{2}}} \frac{\big(\frac{\gamma_Z}{6}t^3 + \frac{\kappa_Z}{24}t^4 + f(t)\big)^2}{t^3}
                  \exp\left\{ \left\vert   \frac{\gamma_Z}{6}t^3 + \frac{\kappa_Z}{24}t^4  + f(t) \right\vert\right\}
              = 1\cdot 0\cdot 1=0,
 \end{align*}
 as desired.
Note that for all \ $k\in\ZZ_+$,
 \begin{align}\label{help2}
  t^k \ee^{\frac{t^2}{2}} = (-1)^k \int_{-\infty}^\infty \ee^{tx}\Phi^{(k+1)}(x)\,\dd x,
  \qquad t\in\RR,
 \end{align}
 where \ $\Phi$ \ denotes the distribution function of a standard normally distributed random variable,
 see, e.g., W\"uthrich  \cite[Lemma 4.4]{Wut}.
Using \eqref{help1} and \eqref{help2} with \ $k=0$, $k=3$ \ and \ $k=4$, \ respectively, we have
 \begin{align}\label{help3}
   \begin{split}
   \int_{-\infty}^\infty \ee^{tx}\,\dd F_Z(x)&= M_Z(t)\\
   &\sim  \int_{-\infty}^\infty \ee^{tx} \left( \Phi^{(1)}(x) - \frac{\gamma_Z}{6} \Phi^{(4)}(x)
              + \frac{\kappa_Z}{24} \Phi^{(5)}(x) \right)\dd x  + o(t^3) \qquad \text{as \ $t\to 0$,}
  \end{split}
 \end{align}
 where \ $F_Z$ \ denotes the cumulative distribution function of \ $Z$.
\ Using that \ $F_Z$ \ is completely determined by its moment generating function \ $M_Z$ \ (see, e.g., W\"uthrich \cite[Lemma 1.2]{Wut})
 and that a counterpart of the continuity theorem holds for random variables having finite moment generating functions
 on an interval containing \ $0$ \ (see, e.g., W\"uthrich  \cite[Lemma 1.4]{Wut}), \eqref{help3} suggests the following approximation
 \[
  \PP(Z\in B) \approx \int_B  \left( \Phi^{(1)}(y) - \frac{\gamma_Z}{6} \Phi^{(4)}(y)
           + \frac{\kappa_Z}{24} \Phi^{(5)}(y) \right)\dd y ,
           \qquad B\in\cB(\RR).
 \]
Here we call the attention to the fact we wrote \ $\approx$ \ and not \ $\sim$,
 \ so, from now on, our approximative formulae are not justified in a rigorous mathematical way;
 they will serve as motivations for introducing new approximations of \ $\VaR$ \ and \ $\ES$ \ of \ $S$ \ in Section \ref{Section_VaR_ES}.
By choosing \ $B=(-\infty,x)$, \ $x\in\RR$, \ we have
 \begin{align}\label{GC}
  F_Z(x) = \PP(Z<x) \approx  \Phi(x) - \frac{\gamma_Z}{6} \Phi^{(3)}(x)
             + \frac{\kappa_Z}{24} \Phi^{(4)}(x)=: \mathrm{GC}_4(x), \qquad x\in\RR.
 \end{align}
The function \ $\mathrm{GC}_4:\RR\to\RR$ \ defined in \eqref{GC} is called a \ $4^{\mathrm{th}}$ order Gram-Charlier type A approximation of \ $F_Z$,
 \ see, e.g., Jondeau et al.\ \cite[formula (5.14)]{JonPooRoc}.
The original NPA method is based on a \ $3^{\mathrm{rd}}$ order Edgeworth approximation of \ $F_Z$ \ given by
 \ $\mathrm{EW}_3(x):= \Phi(x) - \frac{\gamma_Z}{6} \Phi^{(3)}(x)$, \ $x\in\RR$, \ see, e.g., Kaas et al.\ \cite[Remark 2.5.9]{KaaGooDhaDen}.

Using that
 \begin{align}\label{help4}
  \begin{split}
   &\varphi^{(1)}(x)=-x\varphi(x), \qquad x\in\RR,
    \qquad
    \varphi^{(2)}(x)=(x^2-1)\varphi(x), \qquad x\in\RR,\\
   &\varphi^{(3)}(x)= -(x^3-3x)\varphi(x), \qquad x\in\RR,
    \qquad
    \varphi^{(4)}(x)=(x^4-6x^2+3)\varphi(x), \qquad x\in\RR,
   \end{split}
 \end{align}
 where \ $\Phi^{(1)}(x)=\varphi(x)$, \ $x\in\RR$, \ is the standard normal density function,
 we can write \ $\mathrm{GC}_4$ \ in another form, namely,
 \begin{align}\label{form_of_EW_4}
  \mathrm{GC}_4(x)
     = \Phi(x) - \frac{\gamma_Z}{6} \varphi^{(2)}(x) + \frac{\kappa_Z}{24} \varphi^{(3)}(x)
     = \Phi(x) + \left(\frac{\gamma_Z}{6}(-x^2+1) + \frac{\kappa_Z}{24}(-x^3 + 3x) \right) \varphi(x)
 \end{align}
 for \ $ x\in\RR$.
\ Here \ $x$, \ $x^2-1$, \ $x^3- 3x$ \ and \ $x^4-6x^2+3$ \ are the (probabilistic) Hermite polynomials of degree \ $1$, $2$, $3$ \ and \ $4$, \ respectively.

In the next remark we study whether \ $\mathrm{GC}_4$ \ is a distribution function of some random variable or not.

\begin{Rem}\label{Rem_GC}
Note that \ $\mathrm{GC}_4$ \ is continuous, \ $\lim_{x\to-\infty}\mathrm{GC}_4(x)=0$ \ and \ $\lim_{x\to\infty}\mathrm{GC}_4(x)=1$, \
 since \ $\Phi^{(k)}(x) = O(x^{k-1} \ee^{-\frac{x^2}{2}})$ \ as \ $\vert x\vert\to\infty$ \ for all \ $k\geq 2$, $k\in\NN$ \
 (see, e.g., W\"uthrich \cite[Section 4.1.3]{Wut}).
However, we call the attention to the fact that, in general,
 \ $\mathrm{GC}_4$ \ is not a distribution function of some random variable, since, in general,
 \ $\mathrm{GC}_4$ \ may be not non-negative or monotone increasing.
Indeed, by \eqref{help4},
 \begin{align*}
   \mathrm{GC}_4^{(1)}(x)
     & = \varphi(x) - \frac{\gamma_Z}{6}\varphi^{(3)}(x) + \frac{\kappa_Z}{24} \varphi^{(4)}(x)\\
     & = \Big(  \frac{\kappa_Z}{24}x^4 + \frac{\gamma_Z}{6}x^3 - \frac{\kappa_Z}{4}x^2 - \frac{\gamma_Z}{2}x + \frac{\kappa_Z}{8} + 1 \Big)
         \frac{1}{\sqrt{2\pi}} \ee^{-\frac{x^2}{2}}
       =: h(x)\varphi(x),   \qquad x\in\RR.
 \end{align*}
If the polynomial \ $h$ \ (of degree 4 provided that \ $\kappa_Z\ne 0$) \ changes sign, then \ $\mathrm{GC}_4$ \ is not monotone increasing.
But, in case of \ $\kappa_Z>0$, \ we have \ $\lim_{x\to\pm\infty} h(x) = \infty$, \ so there exists an \ $x_0>0$ \ such that
 \ $\mathrm{GC}_4$ \ is monotone increasing on \ $[x_0,\infty)$, \
 and using \eqref{form_of_EW_4} and the fact that \ $\lim_{x\to\infty} x^k \varphi(x) = 0$, $k\in\ZZ_+$,
 \ one can choose \ $x_0$ \ such that \ $\mathrm{GC}_4(x)>0$ \ for all \ $x\geq x_0$.
\ Thus \ $\mathrm{GC}_4$ \ can be used as an approximation of \ $F_Z$ \ in the sense that we approximate \ $F_Z$ \ by
 the function \ $\mathrm{GC}_4$ \ which behaves as a distribution function for large enough \ $x$.
\ This is the same phenomenon as for \ $\mathrm{EW}_3$ \ in case of \ $\gamma_Z>0$, \ see W\"uthrich \cite[Example 4.5]{Wut}.
In case of \ $\kappa_Z<0$, \ we have \ $\lim_{x\to\pm\infty} h(x) = -\infty$, \ so we face up to a problem
 concerning the monotonicity of \ $\mathrm{GC}_4$ \ even for large enough \ $x$, \ similarly
 as for \ $\mathrm{EW}_3$ \ in case of \ $\gamma_Z<0$.
\ One may overcome this difficulty taking into account the fact that most of the loss distributions are skewed to the right
 (i.e., \ $\gamma_Z>0$) \ and have positive excess kurtosis (i.e., \ $\kappa_Z>0$).
\ For example,  if \ $Z$ \ has an exponential distribution with parameter \ $\lambda>0$, \ then
 \ $\gamma_Z=2$ \ and \ $\kappa_Z=6$.
\ If \ $Z$ \ has a Pareto type I distribution with parameters \ $a>4$ \ and \ $c>0$, \ i.e.,
 \[
    F_Z(x) = \PP(Z<x) = \begin{cases}
                            1 - \left(\frac{c}{x} \right)^a & \text{if \ $x\geq c$,}\\
                            0 & \text{if \ $x< c$,}
                         \end{cases}
 \]
 then
 \[
  \gamma_Z = \frac{2(1+a)}{a-3} \sqrt{\frac{a-2}{a}}>0
  \qquad \text{and}\qquad
  \kappa_Z=\frac{6(a^3 + a^2 - 6a -2)}{a(a-3)(a-4)}>0.
 \]
If \ $Z$ \ has a lognormal distribution with parameters \ $\mu\in\RR$ \ and \ $\sigma^2>0$, \ then
 \[
  \gamma_Z = (\ee^{\sigma^2} + 2)\sqrt{\ee^{\sigma^2} - 1}>0
  \qquad \text{and}\qquad
  \kappa_Z = \ee^{4\sigma^2} + 2 \ee^{3\sigma^2} + 3 \ee^{2\sigma^2} - 6 >0.
 \]
If \ $Z$ \ has a compound (Poisson) distribution such that the skewness and excess kurtosis of the claim number and the claim
 severities are positive and finite, then we have \ $\gamma_Z>0$ \ and \ $\kappa_Z>0$, \ see Example \ref{Ex_CP}.
We also point out to the fact that the sign of \ $\gamma_Z$ \ does not play a role in the monotonicity
 of \ $\mathrm{GC}_4$ \ for large values \ $x$.
\proofend
\end{Rem}

In what follows, we derive an extension of the NPA \eqref{NPA}.
For an \ $x\in\RR$, \ we try to find a correction term \ $\delta(x)\in\RR$ \ such that
 \ $F_Z(x+\delta(x)) \approx \Phi(x)$.
\ By \eqref{GC}, we search for a \ $\delta(x)$ \ such that
 \[
   \mathrm{GC}_4(x+\delta(x)) = \Phi(x+\delta(x)) - \frac{\gamma_Z}{6} \Phi^{(3)}(x+\delta(x))
             + \frac{\kappa_Z}{24} \Phi^{(4)}(x+\delta(x)) \approx \Phi(x),
             \qquad x\in\RR.
 \]
For any fixed \ $x\in\RR$, \ let \ $g_x:\RR\to\RR$ \ defined by
 \begin{align}\label{help10_gx}
   g_x(\delta):= \Phi(x) - \Big(\Phi(x+\delta) - \frac{\gamma_Z}{6} \Phi^{(3)}(x+\delta)
             + \frac{\kappa_Z}{24} \Phi^{(4)}(x+\delta) \Big),\qquad \delta\in\RR.
 \end{align}
So our task is to find (or approximate) a root of \ $g_x$, \ where \ $x\in\RR$.
\ We check that if \ $x>\sqrt{3}$, \ $\gamma_Z>0$ \ and \ $\kappa_Z>0$, \ then there exists one positive root of \ $g_x$.
\ It is a consequence of Bolzano's theorem, since \ $g_x$ \ is continuous, \ $g_x(0)>0$ \ and \ $\lim_{\delta\to\infty}g_x(\delta)<0$.
\ Indeed, by \eqref{help4}, we have
  \begin{align}\label{help11}
  \begin{split}
   g_x(0) & = \frac{\gamma_Z}{6} \Phi^{(3)}(x) - \frac{\kappa_Z}{24} \Phi^{(4)}(x)
            = \frac{\gamma_Z}{6} \varphi^{(2)}(x) - \frac{\kappa_Z}{24} \varphi^{(3)}(x) \\
          & = \Big( \frac{\gamma_Z}{6}(x^2-1) - \frac{\kappa_Z}{24} (-x^3+3x) \Big)\varphi(x), \qquad x\in\RR,
  \end{split}
 \end{align}
 so if \ $x>\sqrt{3}$, \ $\gamma_Z>0$ \ and \ $\kappa_Z>0$, \ then \ $g_x(0)>0$.
\ Further, \ $\lim_{\delta\to\infty}g_x(\delta) = \Phi(x) - 1<0$ \ for any \ $x\in\RR$.

Now we turn to derive \eqref{KNPA1}.
We use a $1^{\mathrm{st}}$ order Taylor's approximation of \ $g_x$, \ i.e., \ $g_x(\delta) = g_x(0) +  g_x'(0)\delta + o(\delta)$ \ as
 \ $\delta\to0$ \ (known as ($1^{\mathrm{st}}$  order) Newton's method).
So if \ $g_x(\delta(x))=0$, \ then \ $\delta(x)$ \ can be approximated by \ $-\frac{g_x(0)}{g_x'(0)}$ \
 provided that \ $g_x'(0)\ne 0$, \ where, by \eqref{help4},

 \begin{align}\label{help5}
  \begin{split}
   g_x'(\delta) &= - \Phi^{(1)}(x+\delta) + \frac{\gamma_Z}{6} \Phi^{(4)}(x+\delta) - \frac{\kappa_Z}{24} \Phi^{(5)}(x+\delta)\\
                &= - \varphi(x+\delta) + \frac{\gamma_Z}{6} \varphi^{(3)}(x+\delta) - \frac{\kappa_Z}{24} \varphi^{(4)}(x+\delta), \qquad x,\delta\in\RR,
  \end{split}
 \end{align}
  yielding that
 \begin{align}\label{help12}
  \begin{split}
   g_x'(0) & = - \varphi(x) + \frac{\gamma_Z}{6} \varphi^{(3)}(x) - \frac{\kappa_Z}{24} \varphi^{(4)}(x) \\
           & =  \Big( -1 + \frac{\gamma_Z}{6}(-x^3+3x) - \frac{\kappa_Z}{24} (x^4 - 6x^2 + 3) \Big)\varphi(x), \quad x\in\RR.
  \end{split}
 \end{align}
So
 \begin{align}\label{help6}
  \delta(x) \approx \frac{ - \frac{\gamma_Z}{6}(x^2-1) + \frac{\kappa_Z}{24} (-x^3+3x)}
                        { -1 + \frac{\gamma_Z}{6}(-x^3+3x) - \frac{\kappa_Z}{24} (x^4 - 6x^2 + 3)  },
 \end{align}
 provided that \ $\gamma_Z>0$, \ $\kappa_Z>0$ \ and \ $-1 + \frac{\gamma_Z}{6}(-x^3+3x) - \frac{\kappa_Z}{24} (x^4 - 6x^2 + 3)\ne 0$.
 \ Since \ $\gamma_Z=\gamma_S$ \ and \ $\kappa_Z=\kappa_S$, \ it yields the refinement \eqref{KNPA1} of the NPA \eqref{NPA}.

\begin{Rem}\label{Rem_NR}
(i) Note that the approximation of \ $\delta(x)$ \ given in \eqref{help6} coincides with \ $\delta^{(1)}(x)$, \ where,
   for a given \ $x\in\RR$, \ the sequence \ $(\delta^{(k)}(x))_{k\in\ZZ_+}$ \ is defined via Newton--Raphson's recursion
   \begin{align}\label{NR_recursion}
   \delta^{(k+1)}(x):=\delta^{(k)}(x) - \frac{g_x(\delta^{(k)}(x))}{g_x'(\delta^{(k)}(x))},
     \qquad k\in\ZZ_+,\qquad \delta^{(0)}(x):=0,
   \end{align}
  provided that \ $g_x'>0$, \ where the function \ $g_x$ \ and its derivative \ $g_x'$ \
  are given in \eqref{help10_gx} and \eqref{help5}, respectively.

\noindent (ii) If we formally choose \ $x=z_\alpha$ \ in \ $F_Z(x+\delta(x)) \approx \Phi(x)$, \ where \ $\alpha\in(0,1)$,
 \ then \ $\alpha=\Phi(z_\alpha) \approx F_Z(z_\alpha+\delta(z_\alpha))$, \ so \ $z_\alpha+\delta(z_\alpha)$ \
  is an approximation  of the quantile of \ $Z$ \ at the confidence level \ $\alpha$.
\proofend
\end{Rem}

Next, we derive other refinements of the NPA \eqref{NPA} following the ideas in Remark 2.5.9 in Kaas et al.\ \cite{KaaGooDhaDen}.
If \ $Z$ \ has a compound Poisson distribution with Poisson parameter \ $\lambda>0$ \
 such that the common distribution of the summands (claim severities) has a finite $4^{\mathrm{th}}$-moment,
 then \ $\gamma_Z= O(\lambda^{-1/2})$ \ as \ $\lambda \to\infty$ \ and
 \ $\kappa_Z = O(\lambda^{-1})$ \ as \ $\lambda \to\infty$ \ (see \eqref{OP_moments}),
 which motivate the following refinements of the NPA \eqref{NPA}, similarly
 as in Remark 2.5.9 in  Kaas et al.\ \cite{KaaGooDhaDen}.

If we drop the term for \ $\kappa_Z$ \ in the denominator of the fraction on the right hand side of \eqref{help6},
 then we have
 \[
  \delta(x) \approx  \frac{ - \frac{\gamma_Z}{6}(x^2-1) + \frac{\kappa_Z}{24} (-x^3+3x)}
                          { -1 + \frac{\gamma_Z}{6}(-x^3+3x) },
 \]
 provided that \ $\gamma_Z>0$ \ and \ $-1 + \frac{\gamma_Z}{6}(-x^3+3x) \ne 0$, \ $x\in\RR$, \
 \ yielding the following refinement of the NPA \eqref{NPA}
 \begin{align}\label{KNPA4}
  \PP\left( \frac{S-\EE(S)}{\sqrt{\DD^2(S)}}
            < x + \frac{ - \frac{\gamma_S}{6}(x^2-1) + \frac{\kappa_S}{24} (-x^3+3x)}
                        { -1 + \frac{\gamma_S}{6}(-x^3+3x) } \right)
      \approx \Phi(x),
 \end{align}
 provided that \ $\gamma_S>0$ \ and \ $-1 + \frac{\gamma_S}{6}(-x^3+3x) \ne 0$, \ $x\in\RR$.
\  Note that, if \ $\gamma_S>0$, \ then \ $ -1 + \frac{\gamma_S}{6}(-x^3+3x) <0$
 \ for all \ $x>\sqrt{3}$.

If we drop the term for \ $\kappa_Z$ \ in the numerator of the fraction on the right hand side
 of \eqref{help6}, then we have
 \[
  \delta(x) \approx \frac{ - \frac{\gamma_Z}{6}(x^2-1) }
                        { -1 + \frac{\gamma_Z}{6}(-x^3+3x) - \frac{\kappa_Z}{24} (x^4 - 6x^2 + 3)  },
 \]
 provided that \ $\gamma_Z>0$, \ $\kappa_Z>0$ \ and \ $-1 + \frac{\gamma_Z}{6}(-x^3+3x) - \frac{\kappa_Z}{24} (x^4 - 6x^2 + 3) \ne 0$, \ $x\in\RR$,
 \ yielding the following refinement of the NPA \eqref{NPA}
 \begin{align}\label{KNPA5}
  \PP\left( \frac{S-\EE(S)}{\sqrt{\DD^2(S)}}
            < x + \frac{ - \frac{\gamma_S}{6}(x^2-1) }
                        { -1 + \frac{\gamma_S}{6}(-x^3+3x) - \frac{\kappa_S}{24} (x^4 - 6x^2 + 3) } \right)
      \approx \Phi(x),
 \end{align}
 provided that \ $\gamma_S>0$, \ $\kappa_S>0$ \ and \ $ -1 + \frac{\gamma_S}{6}(-x^3+3x) - \frac{\kappa_S}{24} (x^4 - 6x^2 + 3) \ne 0$, \ $x\in\RR$.

If we drop the term for \ $\kappa_Z$ \ both in the numerator and denominator of the fraction on the right hand side of \eqref{help6},
 then we have
 \begin{align}\label{delta_Kaas}
  \delta(x) \approx \frac{ - \frac{\gamma_Z}{6}(x^2-1) }
                        { -1 + \frac{\gamma_Z}{6}(-x^3+3x) },
 \end{align}
 provided that \ $\gamma_Z>0$ \ and \ $-1 + \frac{\gamma_Z}{6}(-x^3+3x) \ne 0$, \ $x\in\RR$,
 \ yielding the following refinement of the NPA \eqref{NPA}
 \begin{align}\label{KNPA6}
  \PP\left( \frac{S-\EE(S)}{\sqrt{\DD^2(S)}}
            < x + \frac{ - \frac{\gamma_S}{6}(x^2-1) }
                        { -1 + \frac{\gamma_S}{6}(-x^3+3x) } \right)
      \approx \Phi(x),
 \end{align}
 provided that \ $\gamma_S>0$ \ and \ $ -1 + \frac{\gamma_S}{6}(-x^3+3x)  \ne 0$, \ $x\in\RR$.
\ Note that \eqref{KNPA6} can be considered as a refinement of the original NPA \eqref{NPA}
 in the sense that it contains only the first three moments of \ $S$, \ but
 it measures the effect of the skewness \ $\gamma_S$ \ in another way.
For historical fidelity, we note that the approximation for \ $\delta(x)$ \ given in \eqref{delta_Kaas}
 can be found in Kaas et al.\ \cite[formula (2.72)]{KaaGooDhaDen}, where for deriving the original NPA \eqref{NPA}
 they used a \ $3^{\mathrm{rd}}$ order Edgeworth approximation of \ $F_Z$ \ and dropped the term for
 \ $\gamma_Z$ \ in the denumerator of the fraction in \eqref{KNPA6}.

\section{Approximations of \ $\VaR$ \ and \ $\ES$ \ using kurtosis }\label{Section_VaR_ES}

In this section, motivated by part (ii) of Remark \ref{Rem_NR} and by the refinements \eqref{KNPA1}, \eqref{KNPA4}, \eqref{KNPA5}
 and \eqref{KNPA6} of the NPA \eqref{NPA}, we introduce new approximations of \ $\VaR_S(\alpha)$ \ and \ $\ES_S(\alpha)$, \ $\alpha\in(0,1)$.
\ First, we recall the definitions of \ $\VaR$ \ and \ $\ES$ \ of \ $S$ \ in the case of the distribution function \ $F_S$ \ of \ $S$ \ is continuous,
 where \ $S$ \ is typically a loss in the language of insurance mathematics.

\begin{Def}\label{Def_VaR}
Let \ $S$ \ be a random variable such that its distribution function \ $F_S$ \ is continuous.
The Value at Risk of \ $S$ \ at a level \ $\alpha\in(0,1)$ \ is defined by
 \[
    \VaR_S(\alpha):=\inf\{ y\in\RR : F_S(y) > \alpha\}.
 \]
\end{Def}

Note that \ $\VaR_S(\alpha)$ \ coincides with the quantile of \ $S$ \ at a level \ $\alpha\in(0,1)$.

\begin{Def}\label{Def_ES}
Let \ $S$ \ be a random variable such that its distribution function \ $F_S$ \ is continuous
 and \ $\EE(\max(S,0))<\infty$.
\ The Expected Shortfall (also called Conditional Value at Risk) of \ $S$ \ at a level \ $\alpha\in(0,1)$ \ is defined by
 \[
    \ES_S(\alpha):= \frac{1}{1-\alpha} \EE(S \bbone_{\{ S \geq \VaR_S(\alpha)\}}).
 \]
\end{Def}
We call the attention to the fact that the usual correction term \ $\frac{1}{1-\alpha}\VaR_S(\alpha)( 1-\alpha - \PP(S\geq \VaR_S(\alpha)))$ \
 does not appear in the above definition of \ $\ES_S(\alpha)$, \ since \ $F_S$ \ is continuous.
It is known that under the conditions of Definition \ref{Def_ES} we have
 \begin{align}\label{help13}
  \begin{split}
    \ES_S(\alpha) & = \frac{1}{1-\alpha}\int_\alpha^1 \VaR_S(u)\,\dd u
                     = \EE(S \mid S \geq \VaR_S(\alpha))  \\
                  & = \VaR_S(\alpha) + \frac{1}{1-\alpha} \EE((S - \VaR_S(\alpha))^+)
  \end{split}
 \end{align}
 for each \ $\alpha\in(0,1)$.
\ Here the second expression \ $\EE(S \mid S \geq \VaR_S(\alpha))$ \ for \ $\ES_S(\alpha)$ \ coincides with the so-called
 Tail Value at Risk (or Tail Conditional Expectation) of \ $S$ \ at a level \ $\alpha\in(0,1)$, \ since \ $F_S$ \ is continuous in our case.
We also call the attention to the fact that in Casta\~{n}er et al.\ \cite[Definition 3]{CasClaMar}
 the expectation \ $\EE((S - \VaR_S(\alpha))^+)$ \ is called the expected shortfall of \ $S$ \ at a level \ $\alpha$,
 \ but in the literature the notion of Expected Shortfall is commonly defined as in Definition \ref{Def_ES}.

The refinement \eqref{KNPA1} of the NPA \eqref{NPA} motivates the introduction of the following approximations of \ $\VaR$ \ and \ $\ES$, \ respectively.

\begin{Def}\label{Def_VaR_ES}
Let \ $S$ \ be a random variable such that \ $\EE(S^4)<\infty$, \ $\DD^2(S)\ne 0$, \ $\gamma_S>0$, \ $\kappa_S>0$ \ and its distribution function
 \ $F_S$ \ is continuous.
Let us define the approximation \ $\widehat{\VaR_S}^{(I)}(\alpha)$ \ of \ $\VaR_S$ \ of \ $S$ \ at a level \ $\alpha\in(0,1)$ \ by
 \begin{align}\label{VAR_approx_mod}
   \widehat{\VaR_S}^{(I)}(\alpha):=\EE(S) + \sqrt{\DD^2(S)} \left( z_\alpha + \frac{ - \frac{\gamma_S}{6}(z_\alpha^2-1) + \frac{\kappa_S}{24} (-z_\alpha^3+3z_\alpha)}
                                                   { -1 + \frac{\gamma_S}{6}(-z_\alpha^3+3z_\alpha) - \frac{\kappa_S}{24} (z_\alpha^4 - 6z_\alpha^2 + 3)  }
                            \right),
 \end{align}
 provided that \ $-1 + \frac{\gamma_S}{6}(-z_\alpha^3+3z_\alpha) - \frac{\kappa_S}{24} (z_\alpha^4 - 6z_\alpha^2 + 3)\ne0$.
\ Let us define the approximation \ $\widehat{\ES_S}^{(I)}(\alpha)$ \ of the \ $\ES_S$ \ of \ $S$ \ at a level \ $\alpha\in(0,1)$ \ by
 \begin{align}\label{ES_approx_mod}
  \begin{split}
           &\widehat{\ES_S}^{(I)}(\alpha)\\
           &\quad:= \EE(S)
             + \frac{\sqrt{\DD^2(S)}}{1-\alpha} \left( \varphi(z_\alpha)
             +  \int_{z_\alpha}^\infty \frac{ - \frac{\gamma_S}{6}(y^2-1) + \frac{\kappa_S}{24} (-y^3+3y)}
                              { -1 + \frac{\gamma_S}{6}(-y^3+3y) - \frac{\kappa_S}{24} (y^4 - 6y^2 + 3) } \,\varphi(y)\,\dd y
                            \right),
  \end{split}
 \end{align}
 provided that the integral in \eqref{ES_approx_mod} is well-defined and finite.
\end{Def}

\begin{Rem}\label{Rem_VaR_region}
\noindent{(i)}
We call the attention to the fact that in Definition \ref{Def_VaR_ES}
 we do not suppose that the moment generating function of \ $S$ \ is finite in an interval around zero,
 however, it was supposed in Section \ref{Section_ENPA} in order to derive
 a $4^{\mathrm{th}}$ order Gram-Charlier type A expansion of \ $F_S$ \ in \eqref{GC}.
The formulae \eqref{VAR_approx_mod} and \eqref{ES_approx_mod} are meant to be the definitions
 of new approximations of \ $\VaR_S(\alpha)$ \ and \ $\ES_S(\alpha)$, \ respectively,
 they are motivated by the refinement \eqref{KNPA1} of the NPA \eqref{NPA}, and they
 can be used even if the above condition on the moment generating function does not hold (for example, in case of lognormal distributions).
Though their behaviour and precisions should be carefully studied for every particular \ $S$.
\ In the forthcoming Proposition \ref{Pro_VaR_ES_aszimptotika} we describe their asymptotic behaviour
 as \ $\alpha\uparrow 1$, \ and in Examples \ref{Ex_Exp}, \ref{Ex_Par} and \ref{Ex_LN} we study the question
 of their precision in case of exponential, Pareto and lognormal distributions, respectively.

\noindent{(ii)}
For a standard normally distributed random variable \ $\xi$, \ we have \ $\ES_\xi(\alpha) = \frac{\varphi(z_\alpha)}{1-\alpha}$, \ $\alpha\in(0,1)$
 \ (see, e.g., Casta\~{n}er et al.\ \cite[Lemma 1]{CasClaMar}), and this quantity appears in \eqref{ES_approx_mod}.

\noindent{(iii)}
The assumptions \ $\gamma_S>0$ \ and \ $\kappa_S>0$ \ hold for many notable loss distributions, see Remark \ref{Rem_GC}.
Further, if \ $\gamma_S>0$ \ and \ $\kappa_S>0$, \
 then \ $-1 + \frac{\gamma_S}{6}(-z_\alpha^3+3z_\alpha) - \frac{\kappa_S}{24} (z_\alpha^4 - 6z_\alpha^2 + 3)<0$ \
 for all \ $\alpha > \Phi(\sqrt{3+\sqrt{6}})\approx 0.990213$, \ yielding that
 our new approximative formula \eqref{VAR_approx_mod} for \ $\VaR(\alpha)$ \ is well-defined at a confidence level
 \ $\alpha$ \ greater than \ $\Phi(\sqrt{3+\sqrt{6}})$.
\ Indeed, \ $z_\alpha(z_\alpha^2 -3)>0$ \ if \ $z_\alpha>\sqrt{3}$, \ and
 \ $z_\alpha^4 - 6z_\alpha^2 + 3>0$ \ if \ $z_\alpha>\sqrt{3+\sqrt{6}}$.
\ Similarly, if \ $\gamma_S>0$ \ and \ $\kappa_S\in(0,4)$, \ then
 \ $-1 + \frac{\gamma_S}{6}(-z_\alpha^3+3z_\alpha) - \frac{\kappa_S}{24} (z_\alpha^4 - 6z_\alpha^2 + 3)<0$ \
  for all \ $\alpha > \Phi(\sqrt{3})\approx 0.9583677$, \ yielding that
 our new approximative formula \eqref{VAR_approx_mod} for \ $\VaR(\alpha)$ \ is well-defined at a confidence level
 \ $\alpha$ \ greater than \ $\Phi(\sqrt{3})$.
\ Indeed, if \ $\kappa_S\in(0,4)$, \ then \ $- \frac{\kappa_S}{24} (z_\alpha^4 - 6z_\alpha^2 + 3)\leq \frac{\kappa_S}{4}$ \ for any \ $\alpha\in(0,1)$, \
 and \ $z_\alpha^3 - 3z_\alpha = z_\alpha (z_\alpha^2-1) > 0$ \  if \ $z_\alpha>\sqrt{3}$.
Note also that it is reasonable to consider such a high confidence level of $\alpha$ due to the Basel Committee's \cite{BasCom} regulations,
 as it was explained in the Introduction.
In what follows, the constants \ $\Phi(\sqrt{3+\sqrt{6}})$ \ and \ $\Phi(\sqrt{3})$ \ will appear several times, so, for abbreviation,
 we introduce the following notations
  \[
    C_1:=\Phi\left(\sqrt{3+\sqrt{6}}\right) \qquad \text{and} \qquad  C_2:=\Phi(\sqrt{3}).
  \]

\noindent{(iv)}
If \ $\gamma_S>0$ \ and \ $\kappa_S>0$, \ then the integral in \eqref{ES_approx_mod} is well-defined and finite
 for all \ $\alpha\in( C_1, 1)$, \ yielding that our new approximative formula \eqref{ES_approx_mod} for \ $\ES(\alpha)$ \
  is well-defined at a level \ $\alpha$ \ greater than \ $C_1 \approx 0.990213$.
Indeed, due to (iii), if \ $\gamma_S>0$ \ and \ $\kappa_S>0$, \ then
 \ $-1 + \frac{\gamma_S}{6}(-y^3+3y) - \frac{\kappa_S}{24} (y^4 - 6y^2 + 3) < 0$ \ for \ $y>\sqrt{3+\sqrt{6}}$, \
 and for large enough \ $\alpha\in(0,1)$, \ the absolute value of the integrand of the integral in \eqref{ES_approx_mod}
 can be bounded by \ $\varphi(y)$, \ and \ $\varphi$ \ is integrable on \ $\RR$ \ being a density function.

\noindent{(v)}
The form of the approximations defined in \eqref{VAR_approx_mod} and  \eqref{ES_approx_mod} are motivated by
 part (ii) of Remark \ref{Rem_NR}, the extended NPA \eqref{KNPA1} and the facts that
 \begin{align*}
  \VaR_S(\alpha) = \EE(S) + \sqrt{\DD^2(S)} \VaR_{ \frac{S-\EE(S)}{\sqrt{\DD^2(S)}} }(\alpha),
  \qquad \alpha\in(0,1),
 \end{align*}
  and
 \[
 \ES_S(\alpha) = \EE(S) + \sqrt{\DD^2(S)} \ES_{\frac{S-\EE(S)}{\sqrt{\DD^2(S)}} }(\alpha),
  \qquad \alpha\in(0,1).
 \]

\noindent{(vi)}
If \ $S$ \ is a random variable satisfying the conditions of Definition \ref{Def_VaR_ES}, then for any \ $a,b\in\RR$, $a\ne 0$,
 \ the random variable \ $aS+b$ \ satisfies these conditions as well, and
 \begin{align*}
   &\widehat{\VaR_{as+b}}^{(I)}(\alpha) = a\widehat{\VaR_S}^{(I)}(\alpha)+b,\\
   &\widehat{\ES_{as+b}}^{(I)}(\alpha) = a\widehat{\ES_S}^{(I)}(\alpha)+b,
 \end{align*}
 for any \ $\alpha\in\big(C_1,1\big)$.
 \proofend
\end{Rem}

Similarly to \eqref{VAR_approx_mod} and \eqref{ES_approx_mod}, one can introduce the approximations \ $\widehat{\VaR_S}^{(II)}(\alpha)$,
 \ $\widehat{\VaR_S}^{(III)}(\alpha)$, \ $\widehat{\VaR_S}^{(IV)}(\alpha)$ \ of \ $\VaR_S(\alpha)$, \ and
 \ $\widehat{\ES_S}^{(II)}(\alpha)$, \ $\widehat{\ES_S}^{(III)}(\alpha)$, \ $\widehat{\ES_S}^{(IV)}(\alpha)$ \ of \ $\ES_S(\alpha)$, \
 motivated by the refinements \eqref{KNPA4}, \eqref{KNPA5} and \eqref{KNPA6} of the NPA \eqref{NPA}, respectively,
 by deleting the term for the excess kurtosis \ $\kappa_S$ \ in the denumerator, in numerator and both in the numerator and denumerator
  in the formula \eqref{VAR_approx_mod} and \eqref{ES_approx_mod}, respectively.
For example, we present these for \ $\widehat{\VaR_S}^{(IV)}(\alpha)$ \ and \ $\widehat{\ES_S}^{(IV)}(\alpha)$.

\begin{Def}\label{Def_VaR_ES_IV}
Let \ $S$ \ be a random variable such that \ $\EE(\vert S\vert^3)<\infty$, \ $\DD^2(S)\ne 0$, \ $\gamma_S>0$ \ and its distribution function
 \ $F_S$ \ is continuous.
Let us define the approximations \ $\widehat{\VaR_S}^{(IV)}(\alpha)$ \ and \ $\widehat{\ES_S}^{(IV)}(\alpha)$
 \ of \ $\VaR_S(\alpha)$ \ and \ $\ES_S(\alpha)$ \ for \ $\alpha\in(0,1)$ \ by
 \begin{align}\label{VAR_approx_mod_IV}
   \widehat{\VaR_S}^{(IV)}(\alpha):=\EE(S) + \sqrt{\DD^2(S)} \left( z_\alpha + \frac{ - \frac{\gamma_S}{6}(z_\alpha^2-1) }
                                                   { -1 + \frac{\gamma_S}{6}(-z_\alpha^3+3z_\alpha)  }
                            \right),
 \end{align}
 provided that \ $-1 + \frac{\gamma_S}{6}(-z_\alpha^3+3z_\alpha) \ne0$, \ and
 \begin{align}\label{ES_approx_mod_IV}
   \widehat{\ES_S}^{(IV)}(\alpha):=\EE(S) + \frac{\sqrt{\DD^2(S)}}{1-\alpha} \left( \varphi(z_\alpha)
                                                    + \int_{z_\alpha}^\infty
                                                 \frac{ - \frac{\gamma_S}{6}(y^2-1) }{ -1 + \frac{\gamma_S}{6}(-y^3+3y) } \,\varphi(y)\,\dd y
                                                 \right),
 \end{align}
 provided that the integral in \eqref{ES_approx_mod_IV} is well-defined and finite, respectively.
\end{Def}

Note that if \ $\gamma_S>0$, \ then \ $ -1 + \frac{\gamma_S}{6}(-z_\alpha^3+3z_\alpha) <0$ \ for all \ $\alpha>C_2\approx 0.9583677$, \
 and the integrand in \eqref{ES_approx_mod_IV} is well-defined and finite for all \ $\alpha> C_2\approx 0.9583677$ \
 (can be checked similarly as in part (iii) of Remark \ref{Rem_VaR_region}.)

\begin{Pro}\label{Pro_VaR_ES_aszimptotika}
Under the conditions of Definition \ref{Def_VaR_ES}, we have
 \begin{align}\label{VaR_aszimp}
   &\widehat{\VaR_S}^{(I)}(\alpha) \sim \EE(S) + \sqrt{\DD^2(S)}\sqrt{-2\ln(1-\alpha)}
    \qquad \text{as \ $\alpha\uparrow 1$,}\\ \label{ES_aszimp}
   &\widehat{\ES_S}^{(I)}(\alpha) \sim \EE(S) + \sqrt{\DD^2(S)}\sqrt{-2\ln(1-\alpha)}
    \qquad \text{as \ $\alpha\uparrow 1$.}
 \end{align}
\end{Pro}

\noindent{\bf Proof.}
First, recall that, as a consequence of Mill's ratio (see, e.g., Pinelis \cite{Pin}), we have
 \begin{align}\label{Mill}
   \lim_{\alpha\uparrow 1} \frac{z_\alpha}{\sqrt{-2\ln(1-\alpha)}} = 1
   \qquad\text{and}\qquad
   \lim_{\alpha\uparrow 1}  \frac{\varphi(z_\alpha)}{(1-\alpha) z_\alpha} =1.
 \end{align}
Since \ $\kappa_S>0$ \ and \ $\lim_{\alpha\uparrow 1}z_\alpha=\infty$, \ we have \ $\widehat{\VaR_S}^{(I)}(\alpha) \sim \EE(S) + \sqrt{\DD^2(S)}z_\alpha$ \
 as \ $\alpha\uparrow 1$, \ and then, by \eqref{Mill}, we have \eqref{VaR_aszimp}.
Using again that \ $\kappa_S>0$, \ $\lim_{\alpha\uparrow 1}z_\alpha=\infty$ \ and \ $\lim_{\alpha\uparrow 1} \varphi(z_\alpha)=0$, \
 an application of L'Hospital's rule yields that
 \ $\widehat{\ES_S}^{(I)}(\alpha) \sim \EE(S) + \frac{\sqrt{\DD^2(S)}}{1-\alpha} \varphi(z_\alpha)$ \ as \ $\alpha\uparrow 1$.
\ Indeed, by L'Hospital's rule
 \begin{align*}
   &\lim_{\alpha\uparrow 1} \frac{1}{\varphi(z_\alpha)}
                           \int_{z_\alpha}^\infty \frac{ - \frac{\gamma_S}{6}(y^2-1) + \frac{\kappa_S}{24} (-y^3+3y)}
                              { -1 + \frac{\gamma_S}{6}(-y^3+3y) - \frac{\kappa_S}{24} (y^4 - 6y^2 + 3) } \,\varphi(y)\,\dd y \\
   &\quad = \lim_{\alpha\uparrow 1} \frac{1}{z_\alpha\varphi(z_\alpha)}
                              \cdot \frac{ - \frac{\gamma_S}{6}(z_\alpha^2-1) + \frac{\kappa_S}{24} (-z_\alpha^3+3z_\alpha)}
                              { -1 + \frac{\gamma_S}{6}(-z_\alpha^3+3z_\alpha) - \frac{\kappa_S}{24} (z_\alpha^4 - 6z_\alpha^2 + 3) } \,\varphi(z_\alpha)  \\
    &\quad  = \lim_{\alpha\uparrow 1}  \frac{ - \frac{\gamma_S}{6}(z_\alpha - z_\alpha^{-1}) + \frac{\kappa_S}{24} (-z_\alpha^2+3)}
                              { -1 + \frac{\gamma_S}{6}(-z_\alpha^3+3z_\alpha) - \frac{\kappa_S}{24} (z_\alpha^4 - 6z_\alpha^2 + 3) }
     =0 .
 \end{align*}
Finally, by \eqref{Mill} we have \eqref{ES_aszimp}.
\proofend

Note that, as a consequence of Proposition \ref{Pro_VaR_ES_aszimptotika}, under the conditions of Definition \ref{Def_VaR_ES},
 \ $\widehat{\VaR_S}^{(I)}(\alpha)$ \ and \ $\widehat{\ES_S}^{(I)}(\alpha)$ \ behave asymptotically in the same way as \ $\alpha\uparrow 1$.

One can think it over that the corresponding versions of \eqref{VaR_aszimp} and \eqref{ES_aszimp} hold for
 the approximations \ $\widehat{\VaR_S}^{(II)}(\alpha)$, \ $\widehat{\VaR_S}^{(III)}(\alpha)$, \ $\widehat{\VaR_S}^{(IV)}(\alpha)$ \ of \ $\VaR_S(\alpha)$, \ and
 \ $\widehat{\ES_S}^{(II)}(\alpha)$, \ $\widehat{\ES_S}^{(III)}(\alpha)$, \ $\widehat{\ES_S}^{(IV)}(\alpha)$ \ of \ $\ES_S(\alpha)$,
 \ respectively.

In what follows, we evaluate the precisions of the approximations \eqref{VAR_approx_mod} and \eqref{ES_approx_mod}
 in case of some notable loss distributions that are very popular in insurance mathematics,
 namely, in case of the exponential, Pareto type I, lognormal and compound (Poisson) distributions.
This part can be considered as a counterpart of Section 3 in Casta\~{n}er et al.\ \cite{CasClaMar}.
In the introduction we shedded some light on the importance of the above mentioned loss distributions
 underpinning our choice.

\begin{Ex}[Exponential distribution]\label{Ex_Exp}
Let \ $S$ \ be an exponentially distributed random variable with parameter \ $\lambda>0$.
\ Then, using that \ $\EE(S^k) = \frac{k!}{\lambda^k}$, \ $k\in\NN$, \ one can easily have
 \begin{align*}
   \EE(S) = \frac{1}{\lambda}, \qquad \DD^2(S) = \frac{1}{\lambda^2},\qquad \gamma_S=2, \qquad \kappa_S=6.
 \end{align*}
Recall that \ $\VaR_S(\alpha) = -\frac{1}{\lambda}\ln(1-\alpha)$, \ $\alpha\in(0,1)$, \ and
 \begin{align*}
   \ES_S(\alpha) = -\frac{1}{\lambda}\ln(1-\alpha) + \frac{1}{\lambda}
                   = \EE(S) + \VaR_S(\alpha), \qquad \alpha\in(0,1),
 \end{align*}
 see, e.g., Casta\~{n}er et al.\ \cite[formulas (10) and (11)]{CasClaMar}.
Hence, by \eqref{VaR_aszimp}, the difference of \ $\VaR_S(\alpha)$ \ and \ $\widehat{\VaR_S}^{(I)}(\alpha)$ \
 satisfies
 \begin{align*}
   \DiffVaR_S^{(I)}(\alpha) & := \VaR_S(\alpha) - \widehat{\VaR_S}^{(I)}(\alpha)
                        \sim-\frac{1}{\lambda}\ln(1-\alpha) - \frac{1}{\lambda} - \frac{1}{\lambda}\sqrt{-2\ln(1-\alpha)}\\
                      & \sim -\frac{1}{\lambda} \ln(1-\alpha) = -\EE(S)\ln(1-\alpha)
                       \qquad \text{as \ $\alpha\uparrow 1$,}
 \end{align*}
 and, especially, \ $\lim_{\alpha\uparrow 1}\DiffVaR_S^{(I)}(\alpha) = \infty$.
\ Further, by \eqref{ES_aszimp}, the difference of \ $\ES_S(\alpha)$ \ and \ $\widehat{\ES_S}^{(I)}(\alpha)$ \ satisfies
 \begin{align*}
   \DiffES_S^{(I)}(\alpha):= \ES_S(\alpha) - \widehat{\ES_S}^{(I)}(\alpha)
                       & \sim -\frac{1}{\lambda} \ln(1-\alpha) = -\EE(S)\ln(1-\alpha)
                       \qquad \text{as \ $\alpha\uparrow 1$,}
 \end{align*}
 and, especially, \ $\lim_{\alpha\uparrow 1}\DiffES_S^{(I)}(\alpha) = \infty$.
\proofend
\end{Ex}

\begin{Ex}[Pareto type I distribution]\label{Ex_Par}
Let \ $S$ \ be a random variable having a Pareto type I distribution with parameters \ $a>4$ \ and \ $c>0$, \ i.e.,
 \[
    F_S(x) = \PP(S<x) = \begin{cases}
                            1 - \left(\frac{c}{x} \right)^a & \text{if \ $x\geq c$,}\\
                            0 & \text{if \ $x< c$.}
                         \end{cases}
 \]
It is known that
 \begin{align*}
  &\EE(S) = \frac{ac}{a-1}, \qquad
   \DD^2(S) = \frac{ac^2}{(a-1)^2 (a-2)},\\
  &\gamma_S = \frac{2(1+a)}{a-3} \sqrt{\frac{a-2}{a}},
   \qquad
  \kappa_S=\frac{6(a^3 + a^2 - 6a -2)}{a(a-3)(a-4)}.
 \end{align*}
Recall that, for each \ $\alpha\in(0,1)$, \ we have \ $\VaR_S(\alpha) = c(1-\alpha)^{-\frac{1}{a}}$ \ and
 \begin{align}\label{help9}
    \ES_S(\alpha) = \frac{ac}{a-1} (1-\alpha)^{-\frac{1}{a}} = \frac{a}{a-1}\VaR_S(\alpha)
                    = \frac{1}{c}\EE(S)\VaR_S(\alpha),
 \end{align}
 see, e.g., Casta\~{n}er et al.\ \cite[Appendix A.2]{CasClaMar}.
Hence, by \eqref{VaR_aszimp}, the difference \ $\DiffVaR_S^{(I)}(\alpha)$ \ of \ $\VaR_S(\alpha)$ \ and \ $\widehat{\VaR_S}^{(I)}(\alpha)$ \
 satisfies
 \begin{align*}
   \DiffVaR_S^{(I)}(\alpha)  \sim c(1-\alpha)^{-\frac{1}{a}} - \EE(S) - \sqrt{\DD^2(S)}\sqrt{-2\ln(1-\alpha)}
                       \sim c(1-\alpha)^{-\frac{1}{a}}
                       \qquad \text{as \ $\alpha\uparrow 1$,}
 \end{align*}
 since, by L'Hospital's rule,
 \[
  \lim_{\alpha\uparrow 1} \frac{\sqrt{-2\ln(1-\alpha)}}{c(1-\alpha)^{-\frac{1}{a}} } = 0.
 \]
Especially, \ $\lim_{\alpha\uparrow 1}\DiffVaR_S^{(I)}(\alpha) = \infty$.
\ Further, by \eqref{ES_aszimp}, the difference \ $\DiffES_S^{(I)}(\alpha)$ \ of \ $\ES_S(\alpha)$ \ and \ $\widehat{\ES_S}^{(I)}(\alpha)$ \ satisfies
 \begin{align*}
   \DiffES_S^{(I)}(\alpha) &\sim \frac{ac}{a-1} (1-\alpha)^{-\frac{1}{a}}  - \EE(S) - \sqrt{\DD^2(S)}\sqrt{-2\ln(1-\alpha)}\\
                     &\sim \frac{ac}{a-1} (1-\alpha)^{-\frac{1}{a}}
                        = \EE(S)(1-\alpha)^{-\frac{1}{a}}
                       \qquad \text{as \ $\alpha\uparrow 1$,}
 \end{align*}
 and, especially, \ $\lim_{\alpha\uparrow 1}\DiffES_S^{(I)}(\alpha) = \infty$.

Note that for all \ $x>c$, \ $\lim_{a\to\infty} \left( 1 - \left(\frac{c}{x}\right)^a\right) = 1$, \ yielding that
 \ $S\distr c$ \ as \ $a\to\infty$, \ where \ $\distr$ \ denotes convergence in distribution.
In accordance with it, we have
 \[
   \lim_{a\to\infty} \VaR_S(\alpha) = \lim_{a\to\infty} \ES_S(\alpha) =  c , \qquad \alpha\in(0,1),
 \]
 and
 \[
   \lim_{a\to\infty} \widehat{\VaR_S}^{(I)}(\alpha) = \lim_{a\to\infty} \widehat{\ES_S}^{(I)}(\alpha) = c,
   \qquad \alpha\in( C_1 ,1).
 \]
\proofend
\end{Ex}

\begin{Ex}[Lognormal distribution]\label{Ex_LN}
Let \ $S$ \ be a random variable having a lognormal distribution with parameters \ $\mu\in\RR$ \ and \ $\sigma^2>0$.
\ It is known that
 \begin{align*}
  &\EE(S)= \ee^{\mu+\frac{\sigma^2}{2}},\qquad \DD^2(S) = (\ee^{\sigma^2} -1)\ee^{2\mu + \sigma^2},\\
  &\gamma_S = (\ee^{\sigma^2} + 2)\sqrt{\ee^{\sigma^2} - 1},
    \qquad
   \kappa_S = \ee^{4\sigma^2} + 2 \ee^{3\sigma^2} + 3 \ee^{2\sigma^2} - 6.
 \end{align*}
Recall that for the moment generating function \ $M_S$ \ of \ $S$, \ we have \ $M_S(t)=\infty$ \ for any \ $t>0$, \ however,
  as it was noted in part (i) of Remark \ref{Rem_VaR_region}, in principle, we can use our approximations defined in this section.
Recall that, for each \ $\alpha\in(0,1)$, \ we have \ $\VaR_S(\alpha) = \ee^{\mu + \sigma z_\alpha}$ \ and
 \[
    \ES_S(\alpha) = \frac{1}{1-\alpha}  \ee^{\mu+\frac{\sigma^2}{2}}(1-\Phi(z_\alpha - \sigma)),
 \]
 see, e.g., Casta\~{n}er et al.\ \cite[Appendix A.3]{CasClaMar}.
Hence, by \eqref{Mill} and \eqref{VaR_aszimp}, the difference \ $\DiffVaR_S^{(I)}(\alpha)$ \ of \ $\VaR_S(\alpha)$ \ and \ $\widehat{\VaR_S}^{(I)}(\alpha)$ \
 satisfies
 \begin{align*}
   \DiffVaR_S^{(I)}(\alpha) \sim \ee^{\mu + \sigma \sqrt{-2\ln(1-\alpha)}} - \EE(S) - \sqrt{\DD^2(S)}\sqrt{-2\ln(1-\alpha)}
                       \sim \ee^{\mu + \sigma \sqrt{-2\ln(1-\alpha)}}
 \end{align*}
 as \ $\alpha\uparrow 1$, \ and, especially, \ $\lim_{\alpha\uparrow 1}\DiffVaR_S^{(I)}(\alpha) = \infty$.
\ Further, by \eqref{Mill} and \eqref{ES_aszimp}, the difference \ $\DiffES_S^{(I)}(\alpha)$ \ of \ $\ES_S(\alpha)$ \
 and \ $\widehat{\ES_S}^{(I)}(\alpha)$ \ satisfies
 \begin{align*}
   \DiffES_S^{(I)}(\alpha) &\sim  \ee^{\mu+\frac{\sigma^2}{2}} \frac{1-\Phi(z_\alpha-\sigma)}{1-\alpha}  - \EE(S) - \sqrt{\DD^2(S)}z_\alpha\\
                           & \sim \ee^{\mu+\frac{\sigma^2}{2}} \frac{1-\Phi(z_\alpha-\sigma)}{1-\alpha}  \qquad \text{as \ $\alpha\uparrow 1$,}
 \end{align*}
 since, by L'Hospital's rule,
 \begin{align*}
   \lim_{\alpha\uparrow 1}  \frac{1-\Phi(z_\alpha-\sigma)}{1-\alpha}
    & = \lim_{\alpha\uparrow 1}\varphi(z_\alpha-\sigma)(\Phi^{-1})'(\alpha)
     =  \lim_{\alpha\uparrow 1} \frac{\varphi(z_\alpha - \sigma)}{\Phi'(\Phi^{-1}(\alpha))}
     =  \lim_{\alpha\uparrow 1} \frac{\varphi(z_\alpha - \sigma)}{\varphi(z_\alpha)} \\
    & =  \lim_{\alpha\uparrow 1}  \ee^{\sigma z_\alpha - \frac{\sigma^2}{2}}
     = \infty,
 \end{align*}
 and, similarly, using also \eqref{Mill} and \eqref{help4},
 \begin{align*}
   \lim_{\alpha\uparrow 1} \frac{(1-\alpha)z_\alpha}{1-\Phi(z_\alpha-\sigma)}
     = \lim_{\alpha\uparrow 1} \frac{\varphi(z_\alpha)}{1-\Phi(z_\alpha-\sigma)}
     = \lim_{\alpha\uparrow 1} \frac{z_\alpha \varphi(z_\alpha)}{\varphi(z_\alpha-\sigma)}
     = \lim_{\alpha\uparrow 1} z_\alpha \ee^{-\sigma z_\alpha + \frac{\sigma^2}{2}}
     =0.
 \end{align*}
Especially, \ $\lim_{\alpha\uparrow 1}\DiffES_S^{(I)}(\alpha) = \infty$.
\proofend
\end{Ex}

\begin{Ex}[Compound (Poisson) distribution]\label{Ex_CP}
Let us suppose that \ $S=\sum_{i=1}^N X_i$, \ where \ $N$ \ is a non-negative integer-valued random variable, and \ $X_i$, \ $i\in\NN$,
 \ are independent, identically distributed positive random variables such that they are independent of \ $N$ \ as well
 with the convention that \ $S$ \ equals \ $0$ \ whenever \ $N=0$.
\ The distribution of \ $S$ \ is known as a compound distribution, and \ $S$ \ can be interpreted as an aggregate loss amount,
 where \ $N$ \ is the number of claims (also called frequency) and \ $X_i$, $i\in\NN$, \ are the (individual) claim severities (losses).
In all what follows we suppose that \ $\EE(N)\ne 0$ \ and \ $\DD^2(X_1)\ne 0$.
\ If \ $\EE(N^4)<\infty$ \ and \ $\EE(X_1^4)<\infty$, \ then it is known that the mean, variance, skewness and excess kurtosis
 of \ $S$ \ take the following forms
 \begin{itemize}
  \item[(i)] $\EE(S) = \EE(N)\EE(X_1)$,
  \item[(ii)] $\DD^2(S) = \EE(N)\DD^2(X_1) + \DD^2(N)(\EE(X_1))^2$,
  \item[(iii)] \[
          \gamma_S = \frac{\gamma_N (\DD^2(N))^{3/2}(\EE(X_1))^3 + 3 \DD^2(N)\EE(X_1)\DD^2(X_1) + \EE(N)\gamma_{X_1}(\DD^2(X_1))^{3/2}}
                        {( \EE(N)\DD^2(X_1) + \DD^2(N)(\EE(X_1))^2 )^{3/2}},
        \]
        where \ $\gamma_N$ \ and \ $\gamma_{X_1}$ \ denotes the skewness of \ $N$ \ and \ $X_1$, \ respectively,
  \item[(iv)] \[
          \kappa_S= \frac{A}{( \EE(N)\DD^2(X_1) + \DD^2(N)(\EE(X_1))^2 )^2} -3,
        \]
        where
         \begin{align*}
           A&:=(\kappa_{X_1} +3)\EE(N)(\DD^2(X_1))^2 + 4\gamma_{X_1}\DD^2(N)(\DD^2(X_1))^{3/2} \EE(X_1) \\
            &\phantom{:=\,}  + 3\big(\DD^2(N)+\EE(N)(\EE(N) - 1)\big)(\DD^2(X_1))^2
                     +(\kappa_N+3)(\DD^2(N))^2(\EE(X_1))^4\\
             &\phantom{:=\,} + 6\big(\gamma_N (\DD^2(N))^{3/2} + \EE(N)\DD^2(N)\big)(\EE(X_1))^2\DD^2(X_1),
          \end{align*}
          and \ $\kappa_N$ \ and \ $\kappa_{X_1}$ \ denotes the excess kurtosis of \ $N$ \ and \ $X_1$, \ respectively,
 \end{itemize}
 see e.g. Charpentier \cite[page 105]{Cha} (for (i), (ii) and (iii)) and Shevchenko \cite[Proposition 2.2]{She} (for (iv)).
Note that if \ $\gamma_N>0$ \ and \ $\gamma_{X_1}>0$, \ then \ $\gamma_S >0$;
 \ and if \ $\gamma_N>0$, \ $\kappa_N>0$, \ $\gamma_{X_1}>0$ \ and \ $\kappa_{X_1}>0$, \ then
  \ $\kappa_S>0$.
\ Indeed, by algebraic transformations, we have
 \[
   \kappa_S = \frac{\tA}{( \EE(N)\DD^2(X_1) + \DD^2(N)(\EE(X_1))^2 )^2},
 \]
 where
 \begin{align*}
 \tA&:= \kappa_{X_1}\EE(N)(\DD^2(X_1))^2 + 4\gamma_{X_1}\DD^2(N)(\DD^2(X_1))^{3/2} \EE(X_1)
         + 3\DD^2(N)(\DD^2(X_1))^2 \\
    &\phantom{:=\,} + 2\EE(N)(\DD^2(X_1))^2 + \kappa_N(\DD^2(N))^2(\EE(X_1))^4
       + 6\gamma_N (\DD^2(N))^{3/2}(\EE(X_1))^2 \DD^2(X_1),
 \end{align*}
 and \ $\tA>0$ \ provided that \ $\gamma_N>0$, \ $\kappa_N>0$, \ $\gamma_{X_1}>0$ \ and \ $\kappa_{X_1}>0$.

According to our knowledge, in general, there is no closed formulae for \ $\VaR_S(\alpha)$ \ and \ $\ES_S(\alpha)$, \ $\alpha\in(0,1)$.

If \ $N$ \ has a Poisson distribution with parameter \ $\lambda>0$, \ then
 the distribution of \ $S$ \ is known as a compound Poisson distribution, and, as a special case of the above formulae,
 in case of \ $\EE(X_1^4)<\infty$, \ we have
 \begin{align}\label{OP_moments}
  \EE(S) = \lambda\EE(X_1),
  \quad \DD^2(S)=\lambda \EE(X_1^2),
  \quad \gamma_S = \frac{\EE(X_1^3)}{ \sqrt{\lambda} (\EE(X_1^2))^{3/2}},
  \quad \kappa_S= \frac{\EE(X_1^4)}{\lambda(\EE(X_1^2))^2} ,
 \end{align}
 provided that \ $\EE(X_1^2)\ne 0$, \ see, e.g., Shevchenko \cite[Example 2.3]{She}.
Note that \eqref{OP_moments} also shows that for a compound Poisson distribution in question, given the claim
 severity distribution, if the Poisson frequency parameter \ $\lambda$ \ is large enough, then \ $\kappa_S\in(0,4)$, \
 so, by part (iii) of Remark \ref{Rem_VaR_region}, our new approximative formula \eqref{VAR_approx_mod} for \ $\VaR$ \
 is well-defined at a confidence level greater than \ $C_2\approx 0.9583677$.

Further, if \ $N$ \ has a Poisson distribution with parameter \ $\lambda>0$ \ and \ $0<\EE(X_1^2)<\infty$, \ then
 \[
   \frac{S - \EE(S)}{\sqrt{\DD^2(S)}}
   = \frac{\sum_{i=1}^N X_i - \lambda\EE(X_1)}{\sqrt{ \lambda \EE(X_1^2)}}
   \distr\cN(0,1)\qquad \text{as \ $\lambda \to\infty$,}
 \]
 see, e.g., Kaas et al.\ \cite[Theorem 3.7.1]{KaaGooDhaDen}.
Hence, since \ $\Phi$ \ is continuous and strictly increasing, in case of \ $F_S$ \ being continuous,
 we have \ $\frac{S - \EE(S)}{\sqrt{\DD^2(S)}}$ \ converges in quantile to \ $\xi$ \ as \ $\lambda\to\infty$,
 \ i.e.,
 \[
  \VaR_{\frac{S - \EE(S)}{\sqrt{\DD^2(S)}}}(\alpha)\to\VaR_{\xi}(\alpha) = z_\alpha \quad  \text{as \ $\lambda\to\infty$ \ for each \ $\alpha\in(0,1)$,}
 \]
 where \ $\xi$ \ is a standard normally distributed random variable (see Shorack and Wellner \cite[Proposition 5 on page 8 and Exercise 5 on page 10]{ShoWel}),
 yielding that
 \[
    \frac{\VaR_S(\alpha) - \EE(S)}{\sqrt{\DD^2(S)}}\to\VaR_\xi(\alpha) = z_\alpha \quad  \text{as \ $\lambda\to\infty$ \ for each \ $\alpha\in(0,1)$.}
 \]
So it is reasonable to expect that for each \ $\alpha\in(0,1)$, \
 \ $\widehat{\VaR_S}^{(I)}(\alpha)$ \ is close to \ $\EE(S) + \sqrt{\DD^2(S)} \VaR_\xi(\alpha)$ \
 for large values of \ $\lambda$, \ and \ $\widehat{\ES_S}^{(I)}(\alpha)$ \ is close to \ $\EE(S) + \sqrt{\DD^2(S)} \ES_\xi(\alpha)$ \
 for large values of \ $\lambda$.
In fact, if \ $\gamma_S>0$ \ and \ $\kappa_S>0$, \ using \eqref{OP_moments}, we have
 \begin{align*}
   &\lim_{\lambda\to\infty} \frac{\widehat{\VaR_S}^{(I)}(\alpha) - \EE(S)}{\sqrt{\DD^2(S)}}
        = z_\alpha = \VaR_\xi(\alpha), \qquad \alpha\in( C_1,1),\\
   &\lim_{\lambda\to\infty} \frac{\widehat{\ES_S}^{(I)}(\alpha) - \EE(S)}{\sqrt{\DD^2(S)}}
        = \frac{\varphi(z_\alpha)}{1-\alpha} = \ES_\xi(\alpha), \qquad  \alpha\in( C_1 ,1).
 \end{align*}
As a consequence, if \ $\gamma_S>0$ \ and \ $\kappa_S>0$, \ then
 \begin{align}\label{CP_precision}
   \lim_{\lambda\to\infty} \frac{\DiffVaR_S^{(I)}(\alpha)}{\sqrt{\DD^2(S)}}=0,
   \qquad  \alpha\in(C_1,1).
  \end{align}
\proofend
\end{Ex}

\section{Comparison of approximative formulae}\label{Section_comparisons}

In case of some Pareto type I, lognormal and compound Poisson distributions,
 we compare the performance of our new approximative formulae \eqref{VAR_approx_mod} and \eqref{VAR_approx_mod_IV}
 for \ $\VaR(\alpha)$ \ with those of \eqref{VaR_approx_NPA} and \eqref{VaR_approx_CF}
 at a level \ $\alpha$ \ greater than \ $C_1 =  \Phi(\sqrt{3+\sqrt{6}})\approx 0.990213$.
In the introduction we shedded some light on the importance of the above mentioned loss distributions.
Due to Remark \ref{Rem_VaR_region} and the discussion after Definition \ref{Def_VaR_ES_IV},
 for the considered notable loss distributions, the approximative formula \eqref{VAR_approx_mod}
 is well-defined for \ $\alpha> C_1$, \ and the approximative formula \eqref{VAR_approx_mod_IV}
 is well-defined for \ $\alpha>C_2 = \Phi(\sqrt{3})\approx 0.9583677$.
\ In what follows, for simplicity, we will mostly consider the case \ $\alpha>C_1$.

In the following figures we plot the relative error \ $\frac{\DiffVaR_S(\alpha)}{\VaR_S(\alpha)}$, \ $\alpha\in(C_1,1)$, \
 where \ $\DiffVaR_S(\alpha)=\VaR_S(\alpha) - \widehat{\VaR_S}(\alpha)$, \ and \ $\widehat{\VaR_S}(\alpha)$ \ denotes
 the approximation of \ $\VaR_S(\alpha)$ \ using \eqref{VaR_approx_NPA}, \eqref{VaR_approx_CF}, \eqref{VAR_approx_mod} and \eqref{VAR_approx_mod_IV}
 depending on the given figure.
In the figures the abbreviation \ $\DiffVaR/\VaR\%$ \ denotes this quantity in percentage, i.e., multiplied by 100.
For Pareto type I and lognormal distributions \ $S$, \ explicit formulae are available for the theoretical \ $\VaR_S(\alpha)$,
 \ see Examples \ref{Ex_Par} and \ref{Ex_LN}, but for a compound Poisson distribution no such explicit formula is available,
 so we used Monte-Carlo estimate of \ $\VaR_S(\alpha)$ \ with \ $1000000$ \ simulation steps.

For every loss distribution \ $S$ \ we will present two figures,
 on the first figure the left hand side figure corresponds to \eqref{VaR_approx_NPA},
 on the first figure the right hand side figure corresponds to \eqref{VaR_approx_CF},
 on the second figure the left hand side figure corresponds to \eqref{VAR_approx_mod},
 and on the second figure the right hand side figure corresponds to \eqref{VAR_approx_mod_IV}.

We will evaluate these figures from three viewpoints:
 \begin{itemize}
   \item whether the approximative formulae produce a reasonable
         relative error, say less than \ $45\%$ \ in absolute value, or not;
   \item which is the best approximative formula among the four approximative formulae
         \eqref{VaR_approx_NPA}, \eqref{VaR_approx_CF}, \eqref{VAR_approx_mod} and \eqref{VAR_approx_mod_IV}
         that we compare in the sense that the relative error \ $\frac{\DiffVaR_S(\alpha)}{\VaR_S(\alpha)}$ \
         takes the value \ $0$ \ for some \ $\alpha\in(C_1,1)$, \ and if it is the case, then the length
         of the range of the relative error function \ $\frac{\DiffVaR_S(\alpha)}{\VaR_S(\alpha)}$, $\alpha\in(C_1,1)$, \
         is the smallest;
   \item  our new approximative formula \eqref{VAR_approx_mod} which incorporates kurtosis is better or not than
          the known approximative formula \eqref{VaR_approx_CF} which also uses kurtosis (in another way)
          in the sense that \eqref{VAR_approx_mod} produces less relative error in absolute value than \eqref{VaR_approx_CF} or not.
 \end{itemize}

\begin{figure}[h]
 \centering
 \includegraphics[width=5cm,height=5cm]{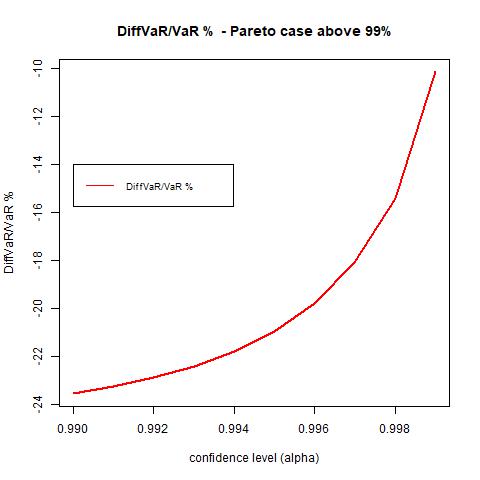}
 \includegraphics[width=5cm,height=5cm]{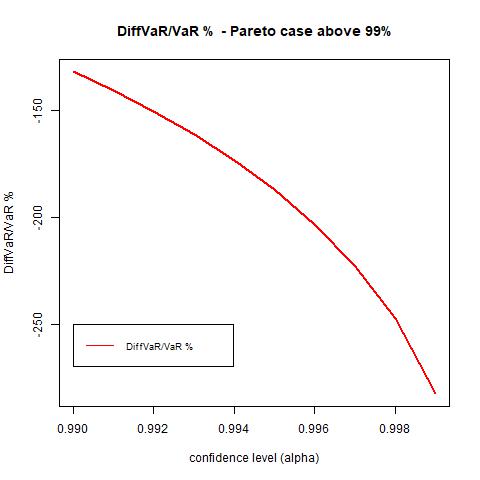}
 \caption{The case of Pareto type I distribution with parameters \ $a=5$ \ and \ $c=10$. \
           On the left figure \ $\VaR$ \ is approximated by \eqref{VaR_approx_NPA}.
           On the right figure \ $\VaR$ \ is approximated by \eqref{VaR_approx_CF}.}
 \label{Fig_Par_NPA_CF}
\end{figure}

\begin{figure}[h]
 \centering
 \includegraphics[width=5cm,height=5cm]{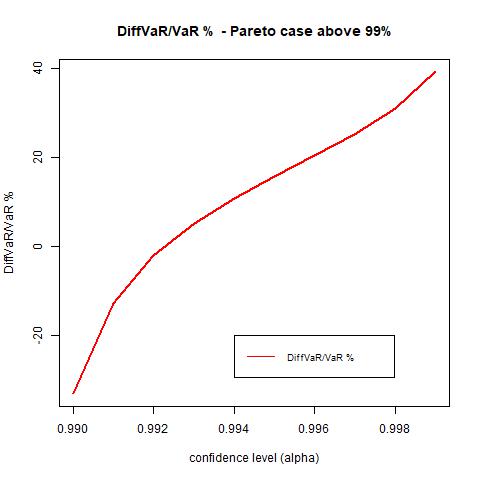}
 \includegraphics[width=5cm,height=5cm]{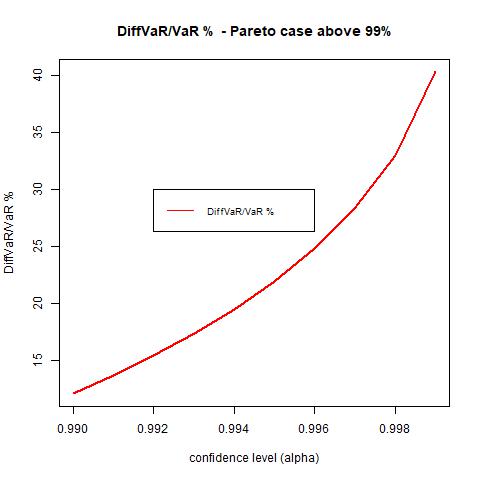}
 \caption{The case of Pareto type I distribution with parameters \ $a=5$ \ and \ $c=10$. \
           On the left figure \ $\VaR$ \ is approximated by \eqref{VAR_approx_mod}.
           On the right figure \ $\VaR$ \ is approximated by \eqref{VAR_approx_mod_IV}.}
 \label{Fig_Par_I_IV}
\end{figure}

\begin{figure}[h]
 \centering
 \includegraphics[width=5cm,height=5cm]{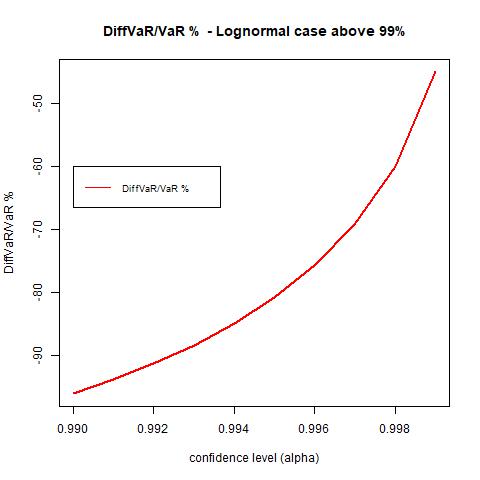}
 \includegraphics[width=5cm,height=5cm]{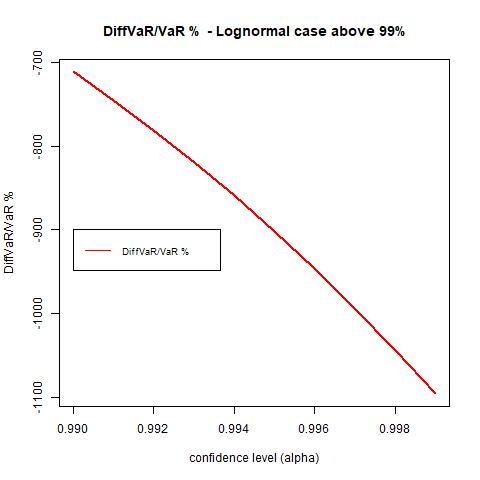}
 \caption{The case of lognormal distribution with parameters \ $\mu=5$ \ and \ $\sigma^2=1.1^2$. \
           On the left figure \ $\VaR$ \ is approximated by \eqref{VaR_approx_NPA}.
           On the right figure \ $\VaR$ \ is approximated by \eqref{VaR_approx_CF}.}
 \label{Fig_LN_NPA_CF}
\end{figure}

\begin{figure}[h]
 \centering
 \includegraphics[width=5cm,height=5cm]{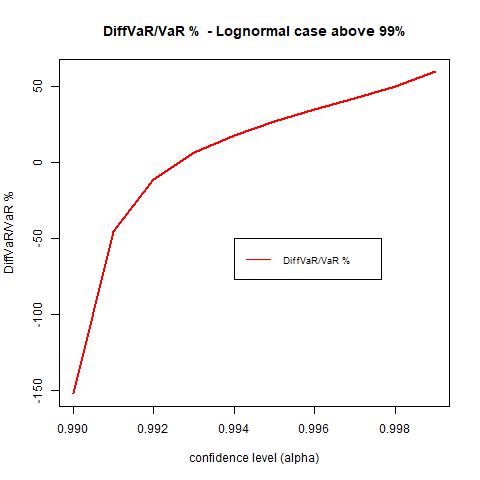}
 \includegraphics[width=5cm,height=5cm]{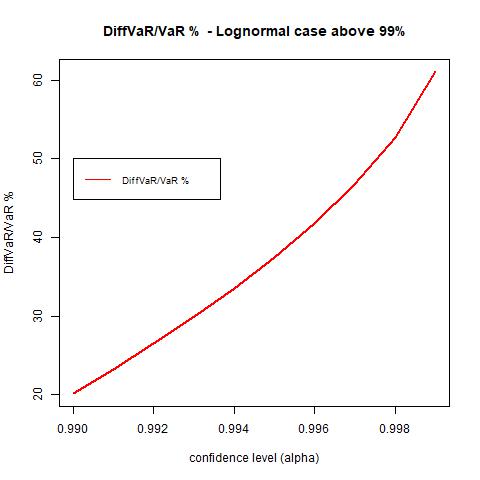}
 \caption{The case of Pareto type I distribution with parameters \ $\mu=5$ \ and \ $\sigma^2=1.1^2$. \
           On the left figure \ $\VaR$ \ is approximated by \eqref{VAR_approx_mod}.
           On the right figure \ $\VaR$ \ is approximated by \eqref{VAR_approx_mod_IV}.}
 \label{Fig_LN_I_IV}
\end{figure}

\begin{figure}[h]
 \centering
 \includegraphics[width=5cm,height=5cm]{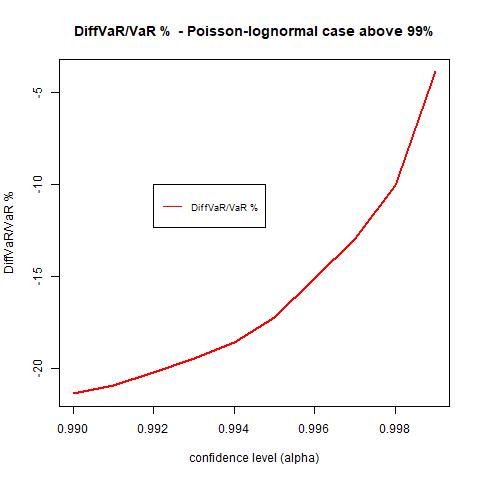}
 \includegraphics[width=5cm,height=5cm]{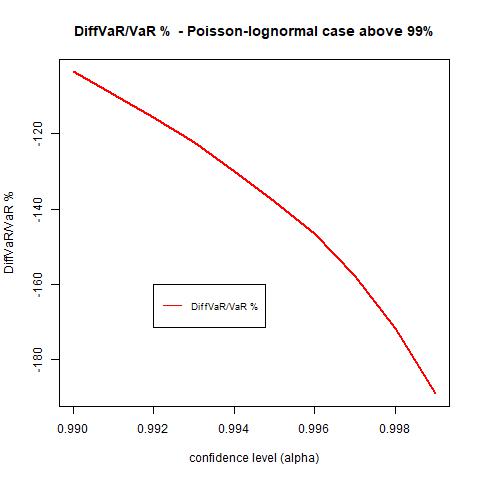}
 \caption{The case of a compound Poisson distribution with frequency parameter \ $\lambda=4$ \ and a lognormal severity distribution  with
           parameters \ $\mu=3$ \ and \ $\sigma^2=1.1^2$.
           On the left figure \ $\VaR$ \ is approximated by \eqref{VaR_approx_NPA}.
           On the right figure \ $\VaR$ \ is approximated by \eqref{VaR_approx_CF}.}
 \label{Fig_CP_NPA_CF}
\end{figure}

\begin{figure}[h]
 \centering
 \includegraphics[width=5cm,height=5cm]{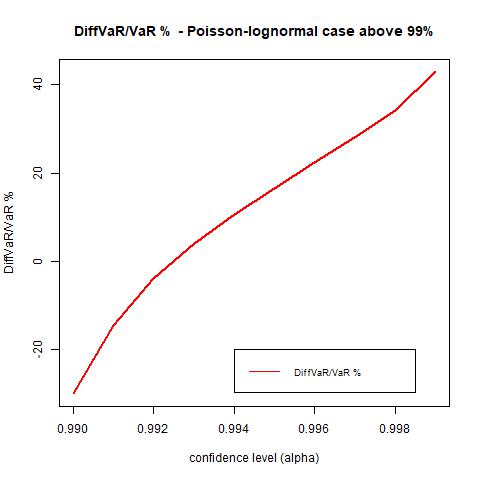}
 \includegraphics[width=5cm,height=5cm]{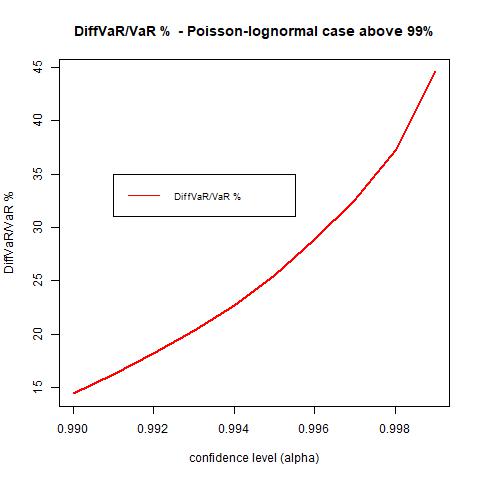}
 \caption{The case of a compound Poisson distribution with frequency parameter \ $\lambda=4$ \ and a lognormal severity distribution with
           parameters \ $\mu=3$ \ and \ $\sigma^2=1.1^2$. \
           On the left figure \ $\VaR$ \ is approximated by \eqref{VAR_approx_mod}.
           On the right figure \ $\VaR$ \ is approximated by \eqref{VAR_approx_mod_IV}.}
 \label{Fig_CP_I_IV}
\end{figure}

For the above three viewpoints, we consider the following loss distributions and give the corresponding figures:
  \begin{itemize}
    \item for \ $S$ \ having a Pareto type I distribution with parameters \ $a=5$ \ and \ $c=10$
      \ (see Example \ref{Ex_Par}), see Figures \ref{Fig_Par_NPA_CF} and \ref{Fig_Par_I_IV},
    \item for \ $S$ \ having a lognormal distribution with parameters \ $\mu=5$ \ and \ $\sigma^2=1.1^2$
      \ (see Example \ref{Ex_LN}), see Figures \ref{Fig_LN_NPA_CF} and \ref{Fig_LN_I_IV},
    \item for \ $S$ \ having a compound Poisson distribution with frequency parameter \ $\lambda=4$ \
          and a lognormal severity distribution with parameters \ $\mu=3$ \ and \ $\sigma^2=1.1^2$ \ (see Example \ref{Ex_CP}),
          see Figures \ref{Fig_CP_NPA_CF} and \ref{Fig_CP_I_IV}.
  \end{itemize}

In Table \ref{Table_1} we summarized the range of \ $\frac{\DiffVaR_S(\alpha)}{\VaR_S(\alpha)}100$, \ $\alpha\in(C_1, 1)$.

\begin{table}[h]
\centering
\begin{tabular}{|c|c|c|c|c|}
  \hline
    $\DiffVaR/\VaR\%$ & \eqref{VaR_approx_NPA} &  \eqref{VaR_approx_CF} & \eqref{VAR_approx_mod} & \eqref{VAR_approx_mod_IV}\\
  \hline
   Pareto type I & $(-24, -10)$ & $(-300, -130)$ & $(-30, 40)$ & $(10, 40)$  \\
   Lognormal & $(-95, -55)$ & $(-1100,-700)$ & $(-150, 50)$ & $(20, 60)$ \\
   Compound Poisson  & $(-20, -5)$ & $(-190,-110)$ & $(-10,40)$ & $(15, 45)$ \\
  \hline
\end{tabular}
\caption{The range of \ $\frac{\DiffVaR_S(\alpha)}{\VaR_S(\alpha)}100$, \ $\alpha\in(C_1, 1)$, \ using the approximative formulae
          \eqref{VaR_approx_NPA}, \eqref{VaR_approx_CF}, \eqref{VAR_approx_mod} and \eqref{VAR_approx_mod_IV} plotted in
           Figures \ref{Fig_Par_NPA_CF}, \ref{Fig_Par_I_IV}, \ref{Fig_LN_NPA_CF}, \ref{Fig_LN_I_IV},
           \ref{Fig_CP_NPA_CF} and \ref{Fig_CP_I_IV}.}
\label{Table_1}
\end{table}

For the given Pareto type I and compound Poisson distributions, one can conclude that \eqref{VaR_approx_NPA}, \eqref{VAR_approx_mod} and \eqref{VAR_approx_mod_IV}
 produce a reasonable relative error (less than \ $45\%$ \ in absolute value), \ the best approximative formula is \eqref{VAR_approx_mod} among the
 four investigated approximative formulae in the earlier given sense,
 and our new approximative formula \eqref{VAR_approx_mod} incorporating kurtosis performs much better than
 \eqref{VaR_approx_CF} known from the literature which also incorporates kurtosis (in another way).
From Table \ref{Table_1}, one can also realize that \eqref{VaR_approx_NPA} and \eqref{VaR_approx_CF} always
 overestimate \ $\VaR_S(\alpha)$, \ $\alpha\in(C_1,1)$, \ while \eqref{VAR_approx_mod_IV} always underestimates it,
 whereas \eqref{VAR_approx_mod} can overestimate and underestimate it as well depending on the value of \ $\alpha$.
In risk assessment, underestimation is a much more serious issue (which might mean insolvency in capital calculation) than overestimation
 (which might mean more capital than necessary).
However, one should bear in mind that \eqref{VAR_approx_mod} can have zero relative error as well which
 shows an advantage of application of this approximation method.

For the given lognormal distribution, one can conclude that \eqref{VAR_approx_mod_IV} produces a more or less reasonable relative error (between \ $20\%$ \ and \ $60\%$),
 \ the best approximative formula is \eqref{VAR_approx_mod_IV} among the four investigated approximative formulae
 in the sense that it has the smallest relative error in absolute value (in this case there is
 no value of \ $\alpha$ for which the relative error is zero), and our new approximative formula \eqref{VAR_approx_mod}
 performs much better then the known \eqref{VaR_approx_CF}.

\begin{figure}[h]
 \centering
 \includegraphics[width=5cm,height=5cm]{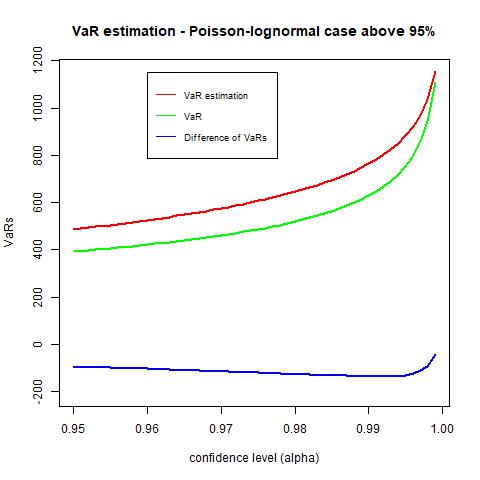}
 \includegraphics[width=5cm,height=5cm]{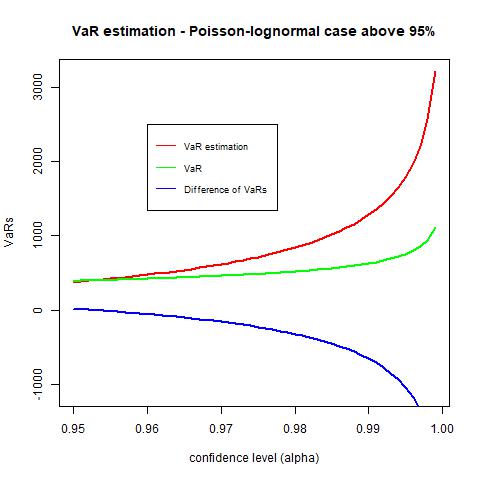}
 \caption{The case of a compound Poisson distribution such that \ $N$ \ has a Poisson
           distribution with frequency parameter \ $\lambda=4$ \ and a lognormal severity distribution with
           parameters \ $\mu=3$ \ and \ $\sigma^2=1.1^2$.
           On the left figure \ $\VaR$ \ is approximated by \eqref{VaR_approx_NPA}.
           On the right figure \ $\VaR$ \ is approximated by \eqref{VaR_approx_CF}.}
 \label{Fig_CP_NPA_CF_csak_VaR}
\end{figure}

\begin{figure}[h]
 \centering
 \includegraphics[width=5cm,height=5cm]{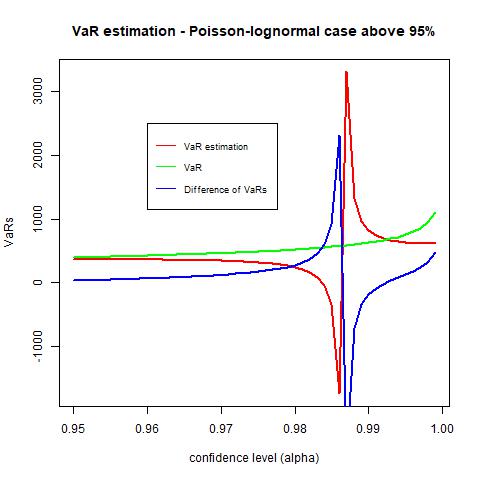}
 \includegraphics[width=5cm,height=5cm]{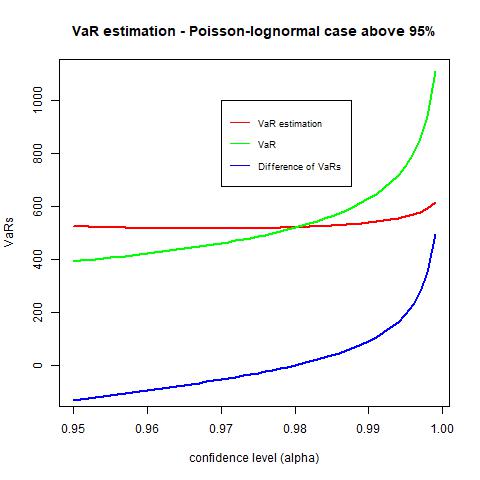}
 \caption{The case of a compound Poisson distribution with frequency parameter \ $\lambda=4$ \ and a lognormal severity distribution with
           parameters \ $\mu=3$ \ and \ $\sigma^2=1.1^2$. \
           On the left figure \ $\VaR$ \ is approximated by \eqref{VAR_approx_mod}.
           On the right figure \ $\VaR$ \ is approximated by \eqref{VAR_approx_mod_IV}.}
 \label{Fig_CP_I_IV_csak_VaR}
\end{figure}

\begin{figure}[h]
 \centering
 \includegraphics[width=5cm,height=5cm]{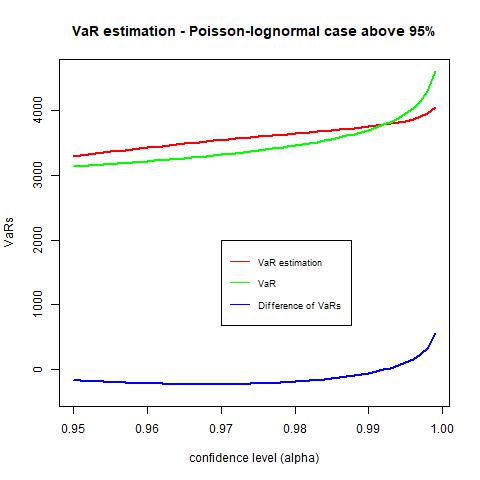}
 \caption{The case of a compound Poisson distribution with frequency parameter \ $\lambda=60$ \ and a lognormal severity distribution with
           parameters \ $\mu=3$ \ and \ $\sigma^2=1.1^2$ \ The \ $\VaR$ \ is approximated by \eqref{VAR_approx_mod}. }
 \label{Fig_CP_I_IV_csak_VaR_l=60}
\end{figure}

Now we turn to presenting the behaviour of our approximations for \ $\VaR$ \ over a larger interval of \ $\alpha$ \ (above \ $0.95$).
We again compare them to those of the ones known from the literature.
In case of \ $S$ \ has a compound Poisson distribution with frequency parameter \ $\lambda=4$ \ and a lognormal severity distribution with
 parameters \ $\mu=3$ \ and \ $\sigma^2=1.1^2$, \ we also plot the functions \ $\VaR_S(\alpha)$, $\alpha\in(0.95,1)$,
 \ $\widehat{\VaR_S}(\alpha)$, $\alpha\in(0.95,1)$, \ and \ $\DiffVaR_S(\alpha)$, \ $\alpha\in(0.95,1)$, \ in one figure
 using the approximative formulae \eqref{VaR_approx_NPA}, \eqref{VaR_approx_CF}, \eqref{VAR_approx_mod} and \eqref{VAR_approx_mod_IV}, respectively,
 see Figures \ref{Fig_CP_NPA_CF_csak_VaR} and \ref{Fig_CP_I_IV_csak_VaR}.
Here, for completeness, we note that the approximative formulae \eqref{VAR_approx_mod} and \eqref{VAR_approx_mod_IV} are well-defined
 on \ $(0.95, C_1]$ \ as well, which explains why we could plotted the functions in question on the interval \ $(0.95, 1)$. \
Remark also that, on Figure \ref{Fig_CP_I_IV_csak_VaR} on left hand side figure, one can see a blow up at around \ $\alpha\approx 0.986567004$, \ which is due to
 the fact that \ $-1 + \frac{\gamma_S}{6}(-z_\alpha^3+3z_\alpha) - \frac{\kappa_S}{24} (z_\alpha^4 - 6z_\alpha^2 + 3)\approx 0$ \ for \ $\alpha\approx 0.986567004$.
\ Note that for the given compound Poisson distribution \ $S$, \ we have \ $\kappa_S\approx 31.61734$, \ and, if one increases the frequency parameter
 \ $\lambda$ \ to \ $60$ \ ($\mu$ \ and \ $\sigma^2$ \ are unchanged), then for the corresponding compound Poisson distribution \ $\tS$ \
 we have \ $\kappa_{\tS}\approx 2.107823 \in (0,4)$,
  \ and in this case, by part (iii) of Remark \ref{Rem_VaR_region},  our new approximative formula \eqref{VAR_approx_mod} for \ $\VaR_{\tS}(\alpha)$ \
  is well-defined at a confidence level \ $\alpha$ \ greater than \ $C_2\approx 0.9583677$ \
  (since \ $-1 + \frac{\gamma_{\tS}}{6}(-z_\alpha^3+3z_\alpha) - \frac{\kappa_{\tS}}{24} (z_\alpha^4 - 6z_\alpha^2 + 3)<0$ \ for \ $\alpha> C_2$
  \ due to the facts that \ $\gamma_{\tS}>0$ \ and \ $\kappa_{\tS}\in(0,4)$).
\ On Figure \ref{Fig_CP_I_IV_csak_VaR_l=60}, we plot the functions \ $\VaR_{\tS}(\alpha)$, $\alpha\in(0.95,1)$,
 \ $\widehat{\VaR_{\tS}}(\alpha)$, $\alpha\in(0.95,1)$, \ and \ $\DiffVaR_{\tS}(\alpha)$, \ $\alpha\in(0.95,1)$, \ in one figure
 using the approximative formula \eqref{VAR_approx_mod}, and, as it is expected, no more blow up appears.

We also call the attention to the fact that the known and the presented new approximative formulae for \ $\VaR_S(\alpha)$ \
 could have a huge relative error and they are sensitive to the choices of parameters.
Namely, for a random variable
 \ $S$ \ having a compound Poisson distribution with frequency parameter \ $\lambda=4$ \ and a lognormal severity distribution
  with parameters \ $\mu=3$ \ and \ $\sigma^2=5^2$ \ (see Example \ref{Ex_CP}),
 the order of \ $\frac{\DiffVaR_S(\alpha)}{\VaR_S(\alpha)}100$, \ $\alpha\in(C_1, 1)$, \ using the approximative formulae
 \eqref{VaR_approx_NPA}, \eqref{VaR_approx_CF}, \eqref{VAR_approx_mod} and \eqref{VAR_approx_mod_IV} is given in Table \ref{Table_2}.
Note that none of the considered approximative methods \eqref{VaR_approx_NPA}, \eqref{VaR_approx_CF}, \eqref{VAR_approx_mod}
 and \eqref{VAR_approx_mod_IV} work well in this case, but \eqref{VAR_approx_mod} and \eqref{VAR_approx_mod_IV} perform much better than \eqref{VaR_approx_NPA} and \eqref{VaR_approx_CF} in terms of order of precision.
In this case \ $\EE(S)=21558794$, \ $\DD^2(S) \approx 8.366638\cdot10^{24}$, \ $\gamma_S \approx 9.6608\cdot 10^{15}$, \ $\kappa_S\approx6.720293 \cdot 10^{42}$
 \ and the relative standard deviation \ $\DD^2(S)/\EE(S)$ \ of \ $S$ \ is \ $134168.6$ \ being quite large.

 \begin{table}[h]
\centering
\begin{tabular}{|c|c|c|c|c|}
  \hline
    $\DiffVaR/\VaR\%$ & \eqref{VaR_approx_NPA} &  \eqref{VaR_approx_CF} & \eqref{VAR_approx_mod} & \eqref{VAR_approx_mod_IV}\\
  \hline
      order & $-10^{22}$ &  $-10^{48}$ & $-10^7$ & $-10^7$  \\
  \hline
\end{tabular}
\caption{Order of \ $\frac{\DiffVaR_S(\alpha)}{\VaR_S(\alpha)}100$, \ $\alpha\in(C_1, 1)$, \ using the approximative formulae
          \eqref{VaR_approx_NPA}, \eqref{VaR_approx_CF}, \eqref{VAR_approx_mod} and \eqref{VAR_approx_mod_IV}
          for a compound Poisson distribution \ $S$ \ with frequency parameter \ $\lambda=4$ \ and a lognormal severity distribution
          with parameters \ $\mu=3$ \ and \ $\sigma^2=5^2$.}
\label{Table_2}
\end{table}

We also note that for an exponential distribution with parameter \ $\lambda>0$, \ the approximative formula \eqref{VaR_approx_NPA}
 (based on the original NPA \eqref{NPA}) gives basically the best performance irrespective of the value of \ $\lambda$ \
 among the four investigated approximative formulae.

\begin{figure}[h]
 \centering
 \includegraphics[width=5cm,height=5cm]{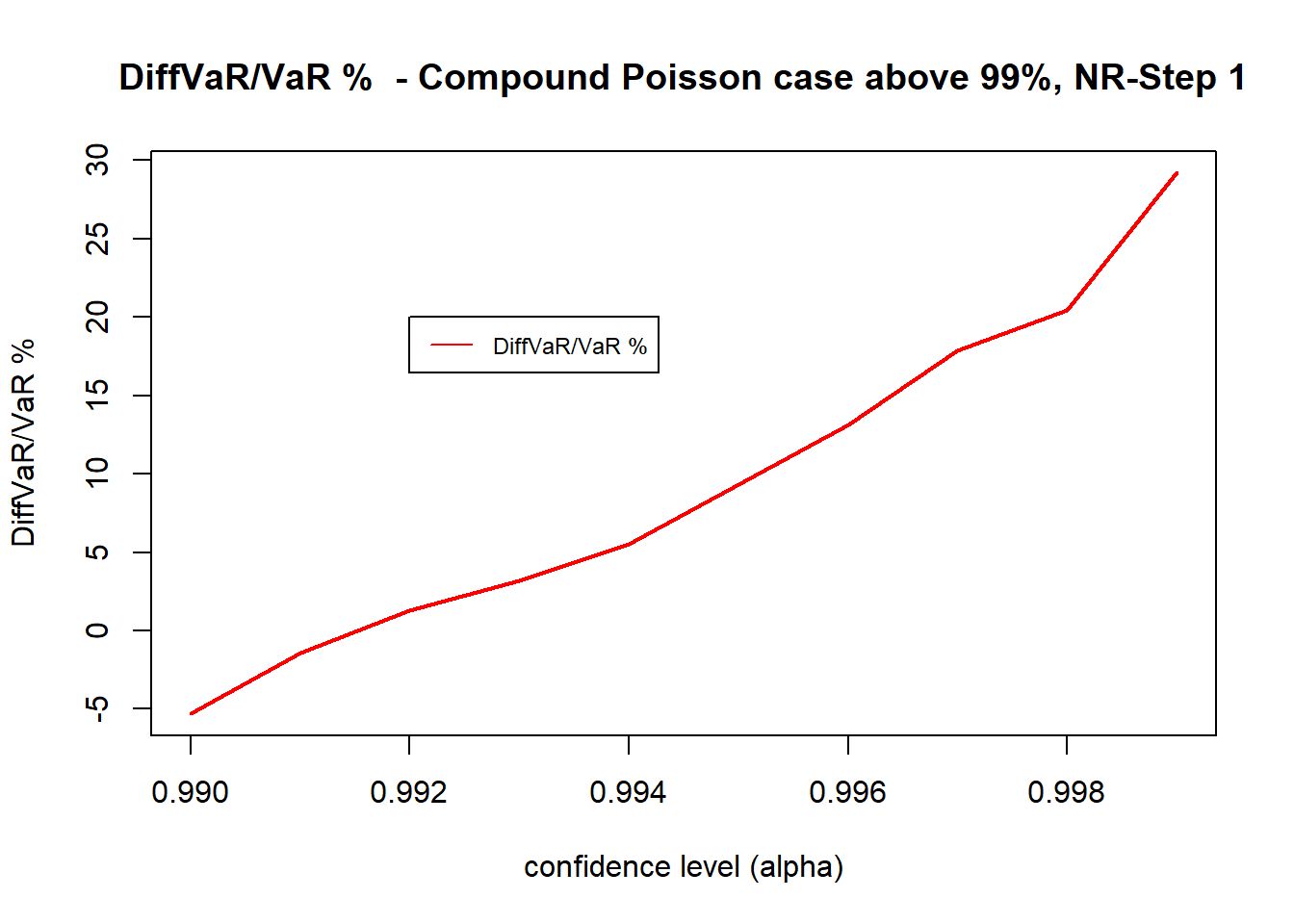}
 \includegraphics[width=5cm,height=5cm]{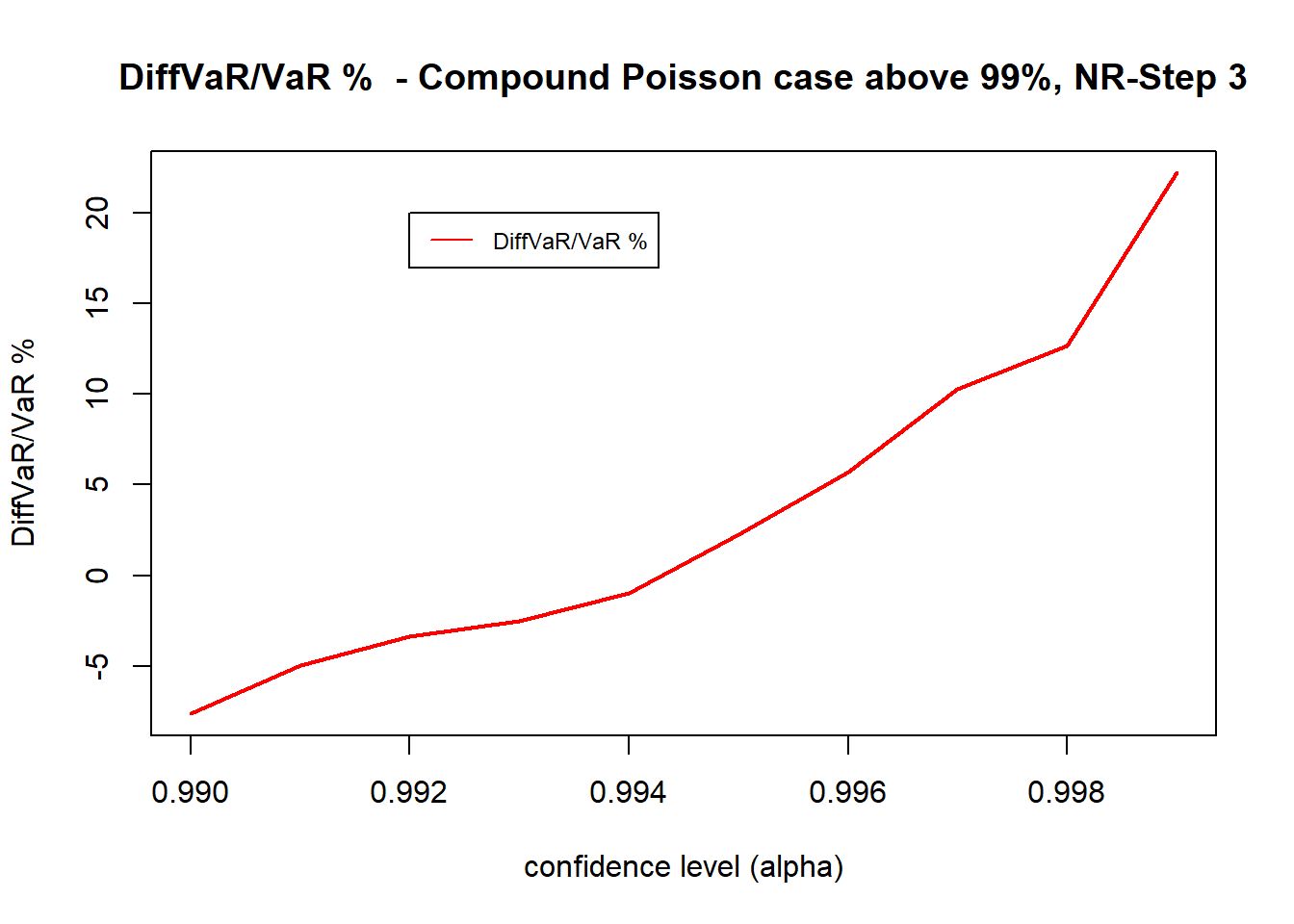}
 \caption{The case of a compound Poisson distribution with frequency parameter \ $\lambda=10$ \ and a lognormal severity distribution with
           parameters \ $\mu=2$ \ and \ $\sigma^2=1$.
           On the left figure \ $\VaR_S(\alpha)$ \ is approximated by \ $\EE(S) + \sqrt{\DD^2(S)}(z_\alpha+ \delta^{(1)}(z_\alpha))$.
           On the right figure \ $\VaR_S(\alpha)$ \ is approximated by \ $\EE(S) + \sqrt{\DD^2(S)}(z_\alpha+ \delta^{(3)}(z_\alpha))$.}
 \label{Figure_NR_1_3}
\end{figure}

\begin{figure}[h]
 \centering
 \includegraphics[width=5cm,height=5cm]{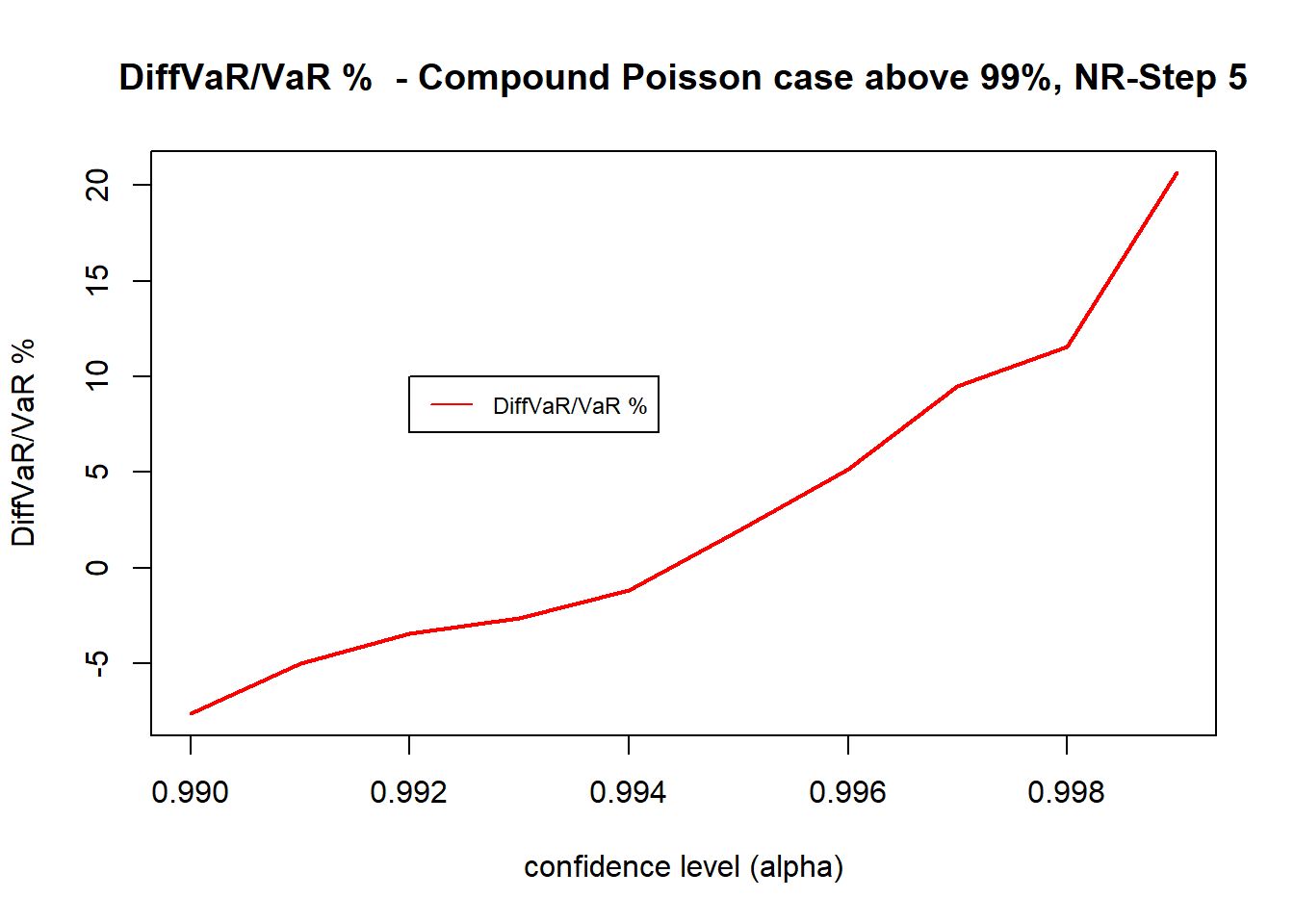}
 \includegraphics[width=5cm,height=5cm]{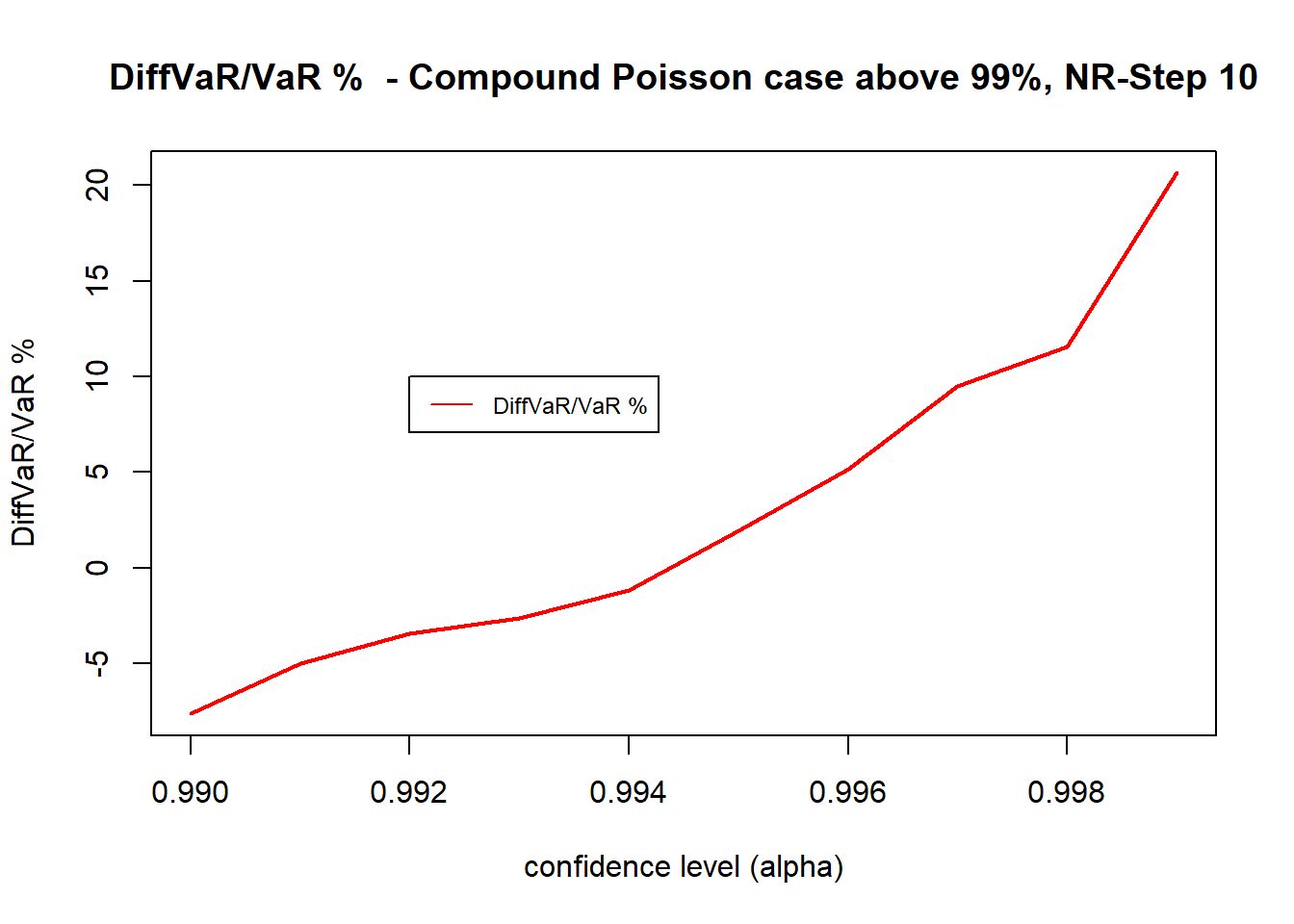}
 \caption{The case of a compound Poisson distribution with frequency parameter \ $\lambda=10$ \ and a lognormal severity distribution with
           parameters \ $\mu=2$ \ and \ $\sigma^2=1$.
           On the left figure \ $\VaR_S(\alpha)$ \ is approximated by \ $\EE(S) + \sqrt{\DD^2(S)}(z_\alpha+ \delta^{(5)}(z_\alpha))$.
           On the right figure \ $\VaR_S(\alpha)$ \ is approximated by \ $\EE(S) + \sqrt{\DD^2(S)}(z_\alpha+ \delta^{(10)}(z_\alpha))$.}
 \label{Figure_NR_5_10}
\end{figure}

Next, we discuss how the Newton--Raphson's method presented in part (i) of Remark \ref{Rem_NR} could improve
 the performance of the approximation of \ $\VaR$.
\ For a random variable \ $S^*$ \ having a compound Poisson distribution with frequency parameter \ $\lambda=10$ \ and a lognormal severity distribution with
 parameters \ $\mu=2$ \ and \ $\sigma^2=1$, \ we plot the relative error \ $\frac{\DiffVaR_{S^*}(\alpha)}{\VaR_{S^*}(\alpha)}$, \ $\alpha\in(C_1,1)$, \
 where this time \ $\widehat{\VaR_{S^*}}(\alpha)$ \ denotes \ $\EE(S^*) + \sqrt{\DD^2(S^*)}(z_\alpha+ \delta^{(k)}(z_\alpha))$ \ with \ $k=1,3,5$ \ and \ $10$,
 \ where \ $\delta^{(k)}(z_\alpha)$, \ $k\in\ZZ_+$, \ is given according to \eqref{NR_recursion}, and for calculating the theoretical \ $\VaR_{S^*}(\alpha)$, \
 we used Monte-Carlo simulation with \ $10000$ \ simulation steps, see Figures \ref{Figure_NR_1_3} and \ref{Figure_NR_5_10}.
Note that, due to part (i) of Remark \ref{Rem_NR}, \ $\EE(S^*) + \sqrt{\DD^2(S^*)}(z_\alpha+ \delta^{(1)}(z_\alpha))$ \ is nothing else but \ $\widehat{\VaR_{S^*}}^{(I)}(\alpha)$.
\ One can realize that the more steps in the Newton--Raphson's recursion \eqref{NR_recursion} are (i.e., the bigger \ $k$ \ is),
 \ the more shrinked the range of the relative error function \ $\frac{\DiffVaR_{S^*}(\alpha)}{\VaR_{S^*}(\alpha)}$, \ $\alpha\in(C_1,1)$ \ is, as we expect.

Concerning the approximative formulae for \ $\ES(\alpha)$, \
 in case of the Pareto type I, lognormal and compound Poisson distributions \ $S$ \ given before,
 we also compared the performance of our new approximative formulae \eqref{ES_approx_mod} and \eqref{ES_approx_mod_IV}
 for \ $\ES(\alpha)$ \ with those of \eqref{ES_approx_NPA} and \eqref{ES_approx_CF}
 at a level \ $\alpha$ \ greater than \ $C_1\approx 0.990213$.
Here we do not present figures, we only note that our new approximative formula \eqref{ES_approx_mod} incorporating kurtosis performs much better then
 the known \eqref{ES_approx_CF} which also incorporates kurtosis (in another way)
 in the sense that \eqref{ES_approx_mod} produces less relative error in absolute value than \eqref{ES_approx_CF}.

\section{Concluding remarks}
In this paper we have derived new approximations for the Value at Risk and the Expected Shortfall at high levels
 of loss distributions with positive skewness and excess kurtosis, and we have described their precisions for notable ones such as for exponential,
 Pareto, lognormal and compound (Poisson) distributions.
Our approximations are motivated by that kind of extensions of the so-called Normal Power Approximation,
 used for approximating the cumulative distribution function of a random variable, which incorporate
 not only the skewness but the kurtosis of the random variable in question as well.
We have shown the performance of our approximations in numerical examples and we have also given comparisons with some known ones in the literature.

We conclude in general that some of our new approximative formulae outperform the other known formulae in many numerical cases.
Namely, in our simulation results, regarding the approximative formulae for \ $\VaR$, \ for the given Pareto type I and compound Poisson distributions,
 \ we have concluded that formula \eqref{VaR_approx_NPA} by Casta\~{n}er et al.\ \cite{CasClaMar},
 and formulae \eqref{VAR_approx_mod} and \eqref{VAR_approx_mod_IV} proposed by us
 produce a reasonable relative error (less than \ $45\%$ \ in absolute value), \ the best approximative formula is our \ $\VaR$ \ approximation
 \eqref{VAR_approx_mod} among the four investigated approximative formulae \eqref{VaR_approx_NPA}, \eqref{VaR_approx_CF},  \eqref{VAR_approx_mod} and \eqref{VAR_approx_mod_IV}
 in the sense that the relative error \ $\frac{\DiffVaR_S(\alpha)}{\VaR_S(\alpha)}$ \ takes the value \ $0$ \
 for some relevant high value of \ $\alpha$ \ ($\alpha\in(C_1,1)$), \ and if it is the case then the length of the range of the function
 \ $\frac{\DiffVaR_S(\alpha)}{\VaR_S(\alpha)}$, $\alpha\in(C_1,1)$, \ is the smallest.
Further, our new approximative formula \eqref{VAR_approx_mod} incorporating kurtosis performs much better then
 formula \eqref{VaR_approx_CF} known from the literature -- which also incorporates kurtosis (in another way) --
 in the sense that \eqref{VAR_approx_mod} produces less relative error in absolute value than \eqref{VaR_approx_CF}.
For the given lognormal distribution, one can conclude that our formula \eqref{VAR_approx_mod_IV} produces
 a more or less reasonable relative error (between \ $20\%$ \ and \ $60\%$), \ the best approximative formula
 is \eqref{VAR_approx_mod_IV} among the four investigated approximative formulae, and our new approximative formula \eqref{VAR_approx_mod}
 performs much better then the known one \eqref{VaR_approx_CF}.

\vspace*{5mm}

\appendix

\vspace*{5mm}

\noindent{\bf\Large Appendix}

\section{First four cumulants of \ $Z$}\label{App_kumulans}

For the random variable \ $Z=(S-\EE(S))/\sqrt{\DD^2(S)}$ \ given in Section \ref{Section_ENPA},
 we determine the cumulants \ $(\log(M_Z))^{(k)}(0)$, \ $k\in\{0,1,2,3,4\}$ \ of \ $Z$.
\ We have
 \[
  (\log(M_Z))^{(0)}(0) = (\log M_Z)(0) = \log(M_Z(0))=\log(1)=0,
 \]
 and
 \begin{align*}
  (\log(M_Z))^{(1)}(t) = \frac{\EE(Z\ee^{tZ})}{M_Z(t)},\qquad t\in(-t_0,t_0),
 \end{align*}
 yielding that
 \[
   (\log(M_Z))^{(1)}(0) = \frac{\EE(Z)}{M_Z(0)} = \EE(Z) = 0.
 \]
We have
 \[
   (\log(M_Z))^{(2)}(t) = \frac{\EE(Z^2\ee^{tZ})M_Z(t) - (\EE(Z\ee^{tZ}))^2 }{(M_Z(t))^2},\qquad t\in(-t_0,t_0),
 \]
 yielding that
 \[
   (\log(M_Z))^{(2)}(0) = \frac{\EE(Z^2) - (\EE(Z))^2}{1^2} = \EE(Z^2) = 1.
 \]
We have
 \[
   (\log(M_Z))^{(3)}(t) = \frac{\EE(Z^3 \ee^{tZ})(M_Z(t))^3 - 3 \EE(Z^2 \ee^{tZ}) \EE(Z\ee^{tZ})(M_Z(t))^2
                                 + 2 (\EE(Z\ee^{tZ}))^3 M_Z(t)}
                               {(M_Z(t))^4}
 \]
 for all \ $t\in(-t_0,t_0)$, \ yielding that
 \[
   (\log(M_Z))^{(3)}(0) = \frac{\EE(Z^3)}{1^4} = \EE(Z^3) = \gamma_Z.
 \]
By some computation, we also have
 \[
   (\log(M_Z))^{(4)}(t) = \frac{A(t)}{(M_Z(t))^8}, \qquad t\in(-t_0,t_0),
 \]
 where
 \begin{align*}
   A(t) &:= \EE(Z^4 \ee^{tZ})(M_Z(t))^7
           -3 (\EE(Z^2 \ee^{tZ}))^2(M_Z(t))^6
           + 2 (\EE(Z \ee^{tZ}))^4 (M_Z(t))^4\\
         &\phantom{:=\,} - 4 \bigg(\EE(Z^3 \ee^{tZ})(M_Z(t))^3
           - 3 \EE(Z^2 \ee^{tZ}) \EE(Z\ee^{tZ})(M_Z(t))^2 \\
         &\phantom{:= - 4 \bigg(\,}
            + 2 (\EE(Z\ee^{tZ}))^3 M_Z(t)\bigg)(M_Z(t))^3 \EE(Z \ee^{tZ}),
 \end{align*}
 yielding that
 \[
   (\log(M_Z))^{(4)}(0) = \frac{\EE(Z^4) - 3 (\EE(Z^2))^2}{1^8} = \EE(Z^4) - 3 = \kappa_Z.
 \]

\end{document}